\documentclass[10pt,letterpaper]{article}
\usepackage[top=0.85in,left=2.75in,footskip=0.75in]{geometry}

\usepackage{amsmath,amssymb}

\usepackage{changepage}

\usepackage[utf8x]{inputenc}

\usepackage{textcomp,marvosym}

\usepackage{cite}

\usepackage{nameref,hyperref}

\usepackage[right]{lineno}

\usepackage{microtype}
\DisableLigatures[f]{encoding = *, family = * }

\usepackage[table]{xcolor}

\usepackage{array}

\usepackage{graphicx,amsmath,amssymb,amsfonts,latexsym,bm,xcolor,booktabs,authblk,endnotes,ifthen}
\usepackage{gensymb}
\usepackage{braket}
\usepackage{subcaption}
\usepackage{geometry}
\usepackage{array}
\usepackage{comment}
\usepackage{wrapfig}
\usepackage{url}
\usepackage{enumitem}

\usepackage[colorinlistoftodos]{todonotes}

\newcolumntype{+}{!{\vrule width 2pt}}

\newlength\savedwidth

\raggedright
\usepackage[aboveskip=1pt,labelfont=bf,labelsep=period,justification=raggedright,singlelinecheck=off]{caption}

\makeatletter
\renewcommand{\@biblabel}[1]{\quad#1.}
\makeatother

\usepackage{lastpage,fancyhdr,graphicx}
\usepackage{epstopdf}
\fancyheadoffset[L]{2.25in}
\fancyfootoffset[L]{2.25in}
\begin{document}
\vspace*{0.2in}

\begin{flushleft}
{\Large
\textbf\newline{{An agent based force vector model of social influence that predicts strong polarization in a connected world}} 
}
\newline
\\
G. Jordan Maclay\textsuperscript{1*\Yinyang}
Moody Ahmad\textsuperscript{2\Yinyang},
\\
\bigskip
\textbf{1} Quantum Fields LLC, St. Charles, IL, USA
\\
\textbf{2} Richland Center, WI, USA
\\

\bigskip

%
%
\Yinyang These authors contributed equally to this work. 





* jordanmaclay@quantumfields.com

\end{flushleft}
\section*{Abstract}
 The model is based on a vector representation of each agent. The components of the vector are the key continuous “attributes” that determine the social behavior of the agent. A simple mathematical force vector model is used to predict the effect of each agent on all other agents. The force law used is motivated by gravitational force laws and electrical force laws for dipoles. It assumes that the force between two agents is proportional to the "similarity of attributes", which is implemented mathematically as the dot product of the vectors representing the attributes of the agents, and the force goes as the inverse square of the difference in attributes, which is expressed as the Euclidean distance in attribute space between the two vectors. The force between the agents may be positive (attractive), zero, or negative (repulsive) depending on whether the angle between the corresponding vectors is less than, equal to, or greater than 90\degree.  A positive force causes the attributes of the agents to become more similar and the corresponding vectors to become more nearly parallel. Interaction between all agents is allowed unless the distance between the attributes representing the agents exceeds a confidence limit (the Attribute Influence Bound) set in the simulation. Agents with similar attributes tend to form groups. For small values of the Attribute Influence Bound, numerous groups remain scattered throughout attribute space at the end of a simulation.  As the Attribute Influence Bound is increased, and agents with increasingly different attributes can communicate, fewer groups remain at the end, and the remaining groups have increasingly different characteristic attributes and approximately equal sizes. With a large Attribute Influence Bound all agents are connected and extreme bi- or tri-polarization results. During the simulations, depending on the initial conditions, collective behaviors of grouping, consensus, fragmentation and polarization are observed as well as certain symmetries specific to the model, for example, the average of the attributes for all agents does not change significantly during a simulation.



\section*{Introduction to Models of Social Influence}
An agent based model of the dissemination of culture between adjacent agents was proposed by Axelrod in 1997 \cite{axelrod}. By culture he meant the set of individual attributes that are subject to social influence. In his model, communication was possible only between adjacent agents that had at least one trait in common.  This approach was based on the assumption that agents with similarity in views were more likely to communicate and to affect each other, and result in local convergence of attitudes.  It also stressed the role of the geographic location of agents. Various modifications to his seminal model have been given \cite{castellano}, for example, adding the possibility of discussion or acceptance of all traits between agents \cite{dypiec}, or adding "cultural drift" that arises from noise-induced transitions \cite{klemm}, or including mass media \cite{rodriguez}. Models for social attitudes have also been proposed that are based on statistical physics \cite{lewenstein}, such as the Ising model based on the behavior of a system of spinning electrons (see \cite{castellano} for a comprehensive review).

Models in which agents are rational, and can update attitudes based on new information are also proposed \cite{neirotti}.  A recent comprehensive review identified three classes of models \cite{flache}: 1) models of assimilative social influence in which agents influence each other to become more similar, leading to consensus; 2) models with similarity biased influence, so-called bounded confidence models, in which only sufficiently similar agents can communicate, and influence each other to become more similar, which may lead to consensus or, if the similarity bias is sufficiently strong, to multiple homogeneous but different opinion clusters; 3) models with repulsive influence, in which agents which are sufficiently dissimilar influence each other to become more different, which may lead to the formation of clusters with maximally opposing views, referred to as bi-polarization, or to what has been termed group polarization. Hybrid models have been developed that combine features of these three classes \cite{lorenz}\cite{duggins}\cite{delvicario}.

Understanding opinion dynamics and social polarization  has become a focus of efforts recently  \cite{brugnoli} \cite{krueger} \cite{hegselmann2} \cite{deffuant}\cite{deffuant2000} \cite{hegselman3}\cite{deffuant}. Polarization has many definitions and nuances \cite{bramson}. By polarization we mean a situation in which a collection of agents becomes divided into a few groups of approximately equal size having very contrasting opinions. It is sometimes referred to as bi-polarization to distinguish from group polarization, which is the tendency for a group to take a position that is more extreme than the initial positions of its members \cite{isenberg}. 

Agent based models have proven to be a powerful tool for theorizing about opinion dynamics \cite{leifeld}.  Understanding the key factors in the evolution of opinion dynamics is one of the primary functions of these models. There are various theories about the source of polarization and consensus in opinion dynamics, including social balance theory that considers dyadic and/or triadic relations in social networks \cite{traag},  or the geometry of the social ties \cite{galam}, mass media communication \cite{mckeown}, authoritarianism\cite{auth}, or some external forces, such as opinion leadership \cite{albi}, or ethnocentric dominance     \cite{hart}.

 There are two basic theories about the impact of new media channels, such as websites and social media, on the social polarization of attributes: 1) that people tend to expose themselves to like minded points of view and avoid dissimilar perspectives, leading to conformity \cite{latane}, and ultimately to more extreme positions and bi-polarization, the echo chamber effect \cite{brugnoli}, which is reinforced by social media algorithms; 2) that new media allow people to encounter more diverse views and thus to have more balanced opinions \cite{lee}. 
Although there is evidence for both these views, some empirical investigations suggest that, in general, people tend to resist facts, holding inaccurate factual beliefs confidently \cite{kuklinsky}, and that corrections frequently fail to reduce misperceptions \cite{nyhan}, and may cause a boomerang or backfire effect \cite{zollo} \cite{mis}.  

Confirmation bias, the tendency to seek, select and interpret information coherently with one's system of beliefs, influences social agents' interactions and the evolution of their attitudes \cite{nickerson}. Two primary cognitive mechanisms have been proposed to explain confirmation bias: 1) challenge avoidance, not wanting to be challenged and possibly being wrong; 2) reinforcement seeking, to find out one is right \cite{brugnoli}. 

In social media, confirmation bias may reflect an agent's belief that it is more likely that individuals with similar views can be trusted than individuals with very different views \cite{mis}. The behavior is implemented in models by the use of a confidence bound, which represents the maximum difference in opinion tolerated between agents that communicate with each other. The boomerang effect has been interpreted as reflecting the lack of trust extended to an agent with very different views  \cite {mis}. 

Different terms have been used to describe repulsive forces: negative influence, rejection, differentiation, reactance. As summarized by Flache et al, this behavior has been linked to balance theory and cognitive dissonance, xenophobia, and social judgment theory \cite{flache}. It has also been proposed that the greater the ego involvement and self relevance, the greater the possibility that negative influences (repulsive forces) result from social interactions with those whose opinions strongly differ \cite{huetanddeff}.

The ideas behind the Force Vector Model, which has the characteristics of a hybrid bounded confidence model with repulsive forces, and its relationship to other models are described in the following section Conceptual Background. Then we present the mathematical structure in Description of the Force Vector Model, followed by Implementing the Force Vector Model in NetLogo, where we discuss details of the model and its operation.  This is followed by Results, where we outline the validation of the model and present the results of a variety of simulations, then a Brief Discussion and a Conclusion.

\section*{Conceptual Background of the Force Vector Model}

The model of Hegselmann and  Krause \cite{hegselmann} for opinion dynamics has been very influential and extensively developed, and we give just a few references  \cite{lorenz} \cite{pluchino}\cite{pineda}\cite{blondel}\cite{hegselmanninsights}.  We assume that the basic HK model may be considered as generally representative of bounded confidence BC models with continuous opinion spaces.  To understand the Force Vector Model it is helpful to be able to compare it to the familiar HK model, for which we give a brief overview. The basic process is computing the opinion of an agent, which is between 0 and 1, at a time interval T in terms of the opinions of all agents in the prior time interval.  The agent's opinion equals the arithmetic mean of the opinions of all agents with which it can communicate in the preceding time interval T-1. Agents with opinions outside the confidence bound $\epsilon$ are not included in the average. 

We assume all opinions are updated simultaneously. In the original "classic" version without bounded confidence, weights were assigned to each interaction in computing the average, but were dropped for simplicity in the BC version \cite{hegselmann}. To illustrate, assume there are three agents with opinions, at time T, of $0.4, 0.6, 0.8$, and $\epsilon= 0.45$, so all agents can communicate.  The first agent A will have an updated opinion of $1.8/3$ or $0.6$.  The same will happen for the other two agents, so consensus happens rapidly in this case.  On the other hand if $\epsilon=0.25$, then $|0.4-0.8|$ would exceed the confidence bound of agent A, and the updated opinion for A would be $0.4+0.6/2=0.5$.  We can interpret the change in the opinion of A as due, in effect, to a force between A and the other agents' opinions. The force causes agent A to modify their opinion, changing it from $0.4$ to $0.5$. The change in A is greater the larger the distance between agent A and the other agent.  In effect we may interpret the BC model as having a force between opinions proportional to the distance between the initial opinions, which is the force law for a spring following Hooke's Law.  This force law means the basic model is linear, but the presence of confidence bounds makes it non-linear.  It is interesting that the force, the effect of one agent on another, increases as the distance between the agents in opinion space increases like a stretched spring. 

Many of the simulations for this BC model have been in 1 dimension although it has been generalized to two dimensions \cite{pluchino}\cite{blondel}\cite{hegselmanninsights}.  In simulations, opinions of agents often change by $15\%$ in one time step, and will often converge on final values in about 20 time steps.  In simulations described by Hegselmann, if the difference in opinion values was less than $10^{-4}$ out of a maximum of 1, then the opinions were grouped into an opinion cluster \cite{hegselmanninsights}.  For plotting in figures, the cluster had a larger range of $10^{-3}$. If the updating for all agents is simultaneous, then all the opinions in a cluster will remain in the same opinion cluster assuming they all have the same confidence bounds \cite{flache}\cite{deffuant}\cite{hegselmanninsights}.

In the basic one dimensional BC model, for small confidence bounds $\epsilon$, about $0.05$, a plurality of seven to ten opinions clusters survive at the end of a simulation that are spaced from each other by roughly $2\epsilon$. For a slightly larger value, $\epsilon$ about $0.2$, roughly symmetric polarization into two groups separated by about $2\epsilon$ is seen. For $\epsilon >0.4$, consensus of all opinions near 0.5 results \cite{hegselmann}. The observation that the intercluster separation averages about $2.2 \epsilon$ has led to the "2R conjecture", where R is the confidence bound, which suggests that this phenomenon of separation by about 2R may be a general one for many multi-agent systems with consensus dynamics \cite{blondel}.

If the confidence interval in the HK model is made asymmetric and dependent on opinions, so it is different for positive deviations than negative deviations, then a variety of asymmetric and polarized behaviors can be simulated, although group polarization is not possible \cite{hegselmann}. In some simulations with polarization, The boundaries at 0 and 1 can play an important role.

It is possible to modify the BC model to obtain repulsive forces if the difference between opinions $O_i$ exceeds a set range \cite{flache}. This can be done by defining the weight for an opinion difference as $\mu(1-\frac{1}{2}|O_{i}(t) -O_{j}(t)|)$ where $\mu$ is a constant convergence factor.  The weight is positive for small differences in opinion and decreases as the difference increases, and goes negative at some point. 

Hegselmann makes some important points about the purpose of the influential HK BC model: it is "in the tradition of radically simplifying models.  It does not aim at a quantitative and exact prediction or explanation of real world phenomena.  Its purpose is more a qualitative understanding of mechanisms in an area where up till now we do not have that much understanding...The BC model is not a realistic model"
\cite{hegselmanninsights}.

We  can explain the Force Vector Model FVM using many of the concepts central to the HK BC model. The goal of the FVM agent-based model is to understand and demonstrate the effect of the mutual influence of an individual agent on other agents, and to faithfully show the development of the entire population of agents as time passes. Because of the different purposes of the FVM and HK models, there are significant differences.  
We treat the simulation as a simulation of a physical system governed by a vector force law, and we expect a continuous evolution of the system in time that does not depend on the details of the coordinate system.  We have used physics as a scaffold and guide, but we have made adjustments to the equations based on the knowledge of human behavior garnered by researchers.  We have a melding of physics, which has proven so capable of describing the behavior of physical systems, and social theory guided by research in human behavior.  The model is focused on the organization of social space, specifically the evolution of attitudes in the social media landscape due to electronic communication \cite{other}.

The FVM with its two dimensional attribute space shows a broad variety of behaviors characterized by consensus, cooperation and polarization, including the formation of groups of agents in different regions in attribute space corresponding to a plurality of opinions or local consensus formation, as well as fragmentation and polarization.  Fragmentation implies splitting into groups of different sizes while polarization implies splitting into groups of comparable sizes. The groups that form correspond to the opinion clusters that appear in the BC model. To date our simulations have been designed to explore the fundamental properties of the model in prototypical cases, with no specific definitions of the attributes of the agents, similar in spirit to Axelrod's presentation of his model \cite{axelrod}.  Ideally the attributes are independent of each other so movement in one direction in the attribute vector space is independent of movement in another direction and one has an orthogonal coordinate system \cite{morse}.

The FVM has a set of parameters and initial conditions that give the model flexibility and, hopefully, allow it to be calibrated and validated against real data. Some of the parameters are linked to physics concepts and some are linked to concepts in social physics. The objective is to provide the adjustable parameters needed to fit a suitable data set, which should help elucidate the relevant dynamics in terms that are known to social scientists. The outcome of a simulation will depend on the initial conditions and on the values of the parameters which are available for a modeler. Whether this leads to correct simulations is not yet known, but we do know this approach has been successful in other areas of endeavor. From the viewpoint of social psychology, these are very complex situations and the patterns of actual behavior probably depend on the particular issues at hand and their relationships.

The FVM model is deterministic like the HK model and many other models. The forces acting on agents depend on their relative locations in attribute space. The forces are simply calculated from the force vector law for all agents that can communicate, and then the agents are all moved accordingly in one time step. If the exact same initial arrangement of agents is used, the results of the simulation are identical, as we have verified. To obtain the exact same arrangement for a normal initial distribution, requires using the same numerical seed for the generation of the random numbers required,  Initial spatial distributions generated by different numerical seeds can produce different results even if the same mean and SD are used.

Each agent A is represented by a vector with components $(x_A, y_A)$, where the components  represent the two attributes that determine the behavior of the agent in the simulation and determine how the agent affects other agents. To date our simulations have been designed to explore the fundamental properties of the model with no specific definitions of the attributes of the agents, similar in spirit to Axelrod's presentation of his model \cite{axelrod}\cite{other}.  The initial placement of agents is such that $-100\le x_A\le100$ and $-100\le y_B\le 100$.  This 200 x 200 area is the visible universe displayed in NetLogo.  We generally choose the initial conditions so most agents remain in this visible region during simulations. However, agents can move beyond this visible area with no limits or boundaries to their movements. Thus the universe for the simulations is unbounded. We chose a visible range of -100 to + 100 to provide more resolution than 0 to 1 without needing decimals, and include negative values to allow vectors to point in all directions from the origin.

The vector for A is drawn from the origin $(0,0)$ to the point $(x_A, y_A)$. A second agent B would be represented by a vector from the origin to the point $(x_B, y_B)$. We quantify the difference in attributes between the two agents by the Euclidean distance $R$ between the agents in attribute space

\begin{equation}
 R = \sqrt{(x_A-x_B)^2 + (y_A - y_B)^2}   .
\end{equation}

We have chosen a force law between agents that varies inversely with the square of the distance R between the agents in attribute space, in a similar way that gravitational forces or electrostatic forces vary with distance. Forces between agents that are well separated are much smaller than forces between agents that are nearby.  This is a significant difference with the HK BC model, where the forces increase with separation.   

The Force Vector Model has a confidence bound called the Attribute Influence Bound ($AIB$). If, for two agents, the distance between attributes, represented by $R$, is greater than $AIB$, $R>AIB$, then the two agents cannot communicate. We have only done simulations with a simple, symmetric $AIB$, with the same value for all agents. The maximum default value of AIB is 280, which is the diagonal distance across the visible universe. (In the NetLogo slider, it can be made as large as desired,)  
 
Agents that can communicate with each other tend to modify the attributes of each other according to the force law. The forces alter the length and direction of the attribute vectors for each agent. At each time step the force on each agent from all other agents  within their $AIB$ is computed. All agents are then updated simultaneously. A typical simulation has thousands of iterations.

One of the most significant differences between hybrid BC models and the Force Vector Model has to do with whether the force between two agents is positive, zero or negative.  To understand this difference we introduce the concept of the \textit{Attribute Orientation} for each agent. The direction of the vector representing the attributes of an agent at any given time is seen as an indication of the Attribute Orientation or attitude of the agent, $\mu \eta \tau \alpha \phi \rho \eta \nu$, meaning the state of and often change in mind or heart.  The AO is generally modified during the simulation as a result of the forces from other agents.  The AO is independent of the length of the vector for the agent. For example, there may be a short vector and a long vector but if both point in the same general direction then the attribute orientations are similar.

The relative direction of the vector for an agent, indicated by the AO, plays a pivotal role in determining how this agent interacts with other agents during the simulation. Consider a hypothetical example to illustrate the effect of the AO: the AO might indicate the agent's political orientation ranging from extremely conservative to extremely liberal.  If the AO indicates a conservative agent, then in a simulation with the FVM the agent would tend to move toward consensus with other agents with the same AO, conservatives, no matter where they fall on the spectrum of conservatism. Similarly the agents with the liberal AOs in the simulation would tend to move toward consensus among themselves. This behavior is reminiscent of the keys on a piano: the note A may be in the base range or in the soprano range, but it is still an A. There may be a final phase in the simulation dominated by repulsive forces in which both groups tend to become increasingly different from each other.

The role of the AO is implemented by the rule that the force between two agents with approximately the same AO is positive, tending to make the vectors more parallel to each other.  Similarly, if the two vectors tend to point in opposite directions or be anti-parallel, and so have opposite AO, then the force between the agents is repulsive and tends to make the vectors more anti-parallel.

In vector language, the sign of the force between two agents is the sign of the dot product of the vectors.   This approach is quite different from the approach in BC models with repulsive forces, where the sign of the force becomes negative if the difference $R$ between the vectors exceeds a bound. Our implementation of negative forces decouples the sign of the force from the $AIB$ and from the separation between agents $R$.

We note that it is not necessary for a model to explicitly include a repulsive force in order for the model to display clustering, fragmentation and polarization. For example, the HK model can show repulsion by use of asymmetric confidence bounds\cite{flache} \cite{hegselman3}. Similarly, the FVM shows these general characteristics without the need for repulsive forces, but the addition of the repulsive forces increases the possible degree of polarization, allows strong group polarization, and increases the complexity of the agent behaviors.  

To understand the formation of groups in the FVM, consider the characteristics of the force law. Our force law predicts that as two similar agents (meaning they have a similar AO) become closer in their attributes, R, the distance between them, is decreasing, and the forces between them, which are causing them to move toward consensus, increase rapidly. In our model, when two similar agents have nearly identical attributes, and $R$ is approaching 0, and the attractive forces between them to reach consensus, which vary as $1/R^2$, become extremely large, approaching infinity as the separation decreases to zero. To avoid infinite forces, when the decreasing $R$ reaches a preset limit, called the Coalescence Radius $CR$, the agents are merged to form a group, akin to the formation of a chemical bond when two ions are brought together. Forces from other agents outside the group are insufficient to overcome the "bond energy" and remove an agent from the group provided the $CR$ is small enough. Similarly in the HK model, with simultaneous updating, an agent cannot leave an opinion cluster.    

For example, for the default Coalescence Radius of 0.2 (1 part out of 1000 in our visible coordinate system), the force between two agents within the $CR$ is at least 100 times greater than the force between one of these agents and an agent that is a distance 2 from the group.  Typically, in this type of arrangement, the nearby agent will join the group. Groups are assumed to obey the same force laws as individual agents.

There are two parameters that characterize each agent that we have not yet described, the \textit{active mass} and the \textit{passive mass}, defined in analogy with gravitational physics.  These parameters scale 1) how large the force is from a particular agent and 2) how much the agent's attributes change in response to the force from another agent.  The force from one agent on another is proportional to the agent's active mass, which might be likened to charisma or influence. The active mass determines how strong the influence or force from this particular agent is on the other agent. On the other hand, the passive mass indicates how much resistance the agent has to being influenced, and might be likened to stubbornness, inflexibility, or contrarian behavior. Ultimately the ratio of the active mass to the passive mass scales the change in attributes due to the force of interaction. To date most of our simulations have had the active and passive masses equal to each other and described by a normal distribution. We have also explored the effect on the simulations of including selected agents with very large active masses, corresponding to highly influential agents.

The FVM follows general guidelines for modeling most physical systems, in contrast to the HK model.  For example, in  general we do not expect meaningful results to depend on characteristics of the particular coordinate system used. Thus in our simulations, we interpret with caution effects that involve properties of the rectangular boundaries since they would, of course, be different if we used polar instead of rectangular coordinates.

It is in this spirit, that if an agent reaches a boundary of the visible 200 x 200 universe, the icon of the agent stays at the boundary at the point of exit, but the force and distance calculations for the agent continue indefinitely. The agent, now termed a ghost, will continue to move under the same force law as before, although its motion will not be visible.  Inspection of a ghost agent reveals its location.  This means that calculated quantities, like the forces, the mean distance between agents, location of agents, etc., will not be affected by the intersection of the agent trajectory with the visible boundaries, so in effect the universe is without bound.

Agents near the border or ghosts on the borders could conceivably be drawn toward the center region, but we have never observed this, and it seems to be contrary to the common evolution of the FVM simulations, in which outward movement is the trend.  Our intent is that the universe of attributes is sufficiently large so that, in general, the population of agents will tend to be in the visible region.  

Simulation of a physical system is generally expected to show a continuous evolution without major discontinuities, unless the system is chaotic or displays certain phase changes. This FVM requirement for near infinitesimal continuity is in contrast to the HK model, where opinions can change significantly in one time interval and simulations may take only 20 intervals \cite{hegselmann}\cite{hegselmanninsights}. In each time step $DT$ of our simulations we ensure that the largest movement of any agent (group or individual) is less than a small preset amount, the Coalescence Radius $CR$, mentioned earlier in regard to group formation.  We have used the default value of 0.2 for the $CR$ in most simulations, which means that it would take 200/0.2 = 1000 steps to cross from one side of the visible attribute space to the other.  The movement of one step, 0.1\% of the width of the attribute space, is imperceptible. Simulations take thousands of iterations.  We have verified that the effort to make the simulations continuous has resulted in a system that is stable with respect to changes in model parameters.

In order to limit the change in attributes in each time step $DT$ to the Coalescence Radius, we use adaptive time stepping. We determine the value of $DT$ needed to ensure that the largest distance moved by any agent is $CR$, the Coalescence Radius. This provides a smooth, continuous simulation, and greatly reduces the time to complete the simulation. The value of $DT$ is indicative of the strength of the force. For example, if very large values of $DT$ are required to provide a motion of $CR$, the forces are correspondingly weak. This typically occurs when the dominant forces are repulsive and the separations are large.  We have verified that our implementation of adaptive time stepping does not change the results of the simulation. 

During a simulation, the location of an individual agent is shown as a circle, with a number next to it indicating the mass.  Different colors for agents are assigned randomly by NetLogo.  The movement of the agent is indicated by a track of the same color. When agents merge and groups are formed, indicated by a triangle, they take the color and location of the most massive participant in the formation of the group. The size of the triangle is proportional to the log of the mass.  

The origin (0,0) is indicated by a small bullseye.  During the simulation, numerous quantities are monitored, such as the number of groups, the mean distance between all agents, and the mean distance of all agents to the origin.  One of the most important metrics is the mean distance between an agent and the agent closest to it, which we describe as the neighbor, thus we define MDCN = mean distance to closest neighbor. 

For FVM simulations with normal initial spatial distributions in x and y of agents with the same SD, centered about the origin, regularities are observed at the end of simulations: 1) the number of agents remaining decreases approximately exponentially with the $AIB$. Short range attractive forces are primarily responsible for the merging of nearby agents into groups, which reduces the number of agents; 2) for AIB$<50$ many groups are spaced around attribute space, a representation of local consensus formation seen in other models \cite{flache}; 3) for $AIB$ over 70, only two or three groups of comparable mass remain in the simulation and 4) the distance between them is approximately AIB, as they are driven to opposite sides of the origin by repulsive forces; 5) the Center of Attributes, $(X_C,Y_C)$,  the mass weighted mean value of attributes for all agents, whose location is indicated by a small colored rectangle during the simulation, is approximately constant.

Two representative simulations, which will be discussed in Results, can be viewed at \url{https://youtu.be/WzblcRCjiK8} and \url{https://youtu.be/BArqDp8-JAQ}

Our challenge, the challenge of many who work hard to simulate social influence, is to develop "collaborations with empirical researchers where possible, in order to improve the match of empirical evidence and model assumptions" \cite{flache}.


\section*{Description of the Force Vector Model of mutual influence}

The FVM model, programmed in NetLogo, is based on a vector representation of the agents. The model code is available at \url{http://modelingcommons.org/browse/one_model/6747}. 
The key aspects of the model are described in the section below.

\subsection*{The force calculation}
We characterize an agent A by two numbers $(x_A,y_A)$ corresponding to two different attributes. We represent an agent by a point $(x_A,y_A)$ in a two dimensional Cartesian coordinate system (attribute space) or equivalently by the vector $\bm{A}(x_A,y_A)$ from the origin (0,0) to the point $(x_A,y_A)$. The 200 x 200 region where $-100 \le x_A \le 100$ and $-100\leq y_A \leq +100$ represents the visible region of attribute space where agents are initially located. During simulations agents may move beyond this visible space without limit and all force and movement calculations continue. The agent vector represents the significant attributes of the agent for the simulation.  \medskip
\begin{figure}[h]
\includegraphics[scale=0.62]{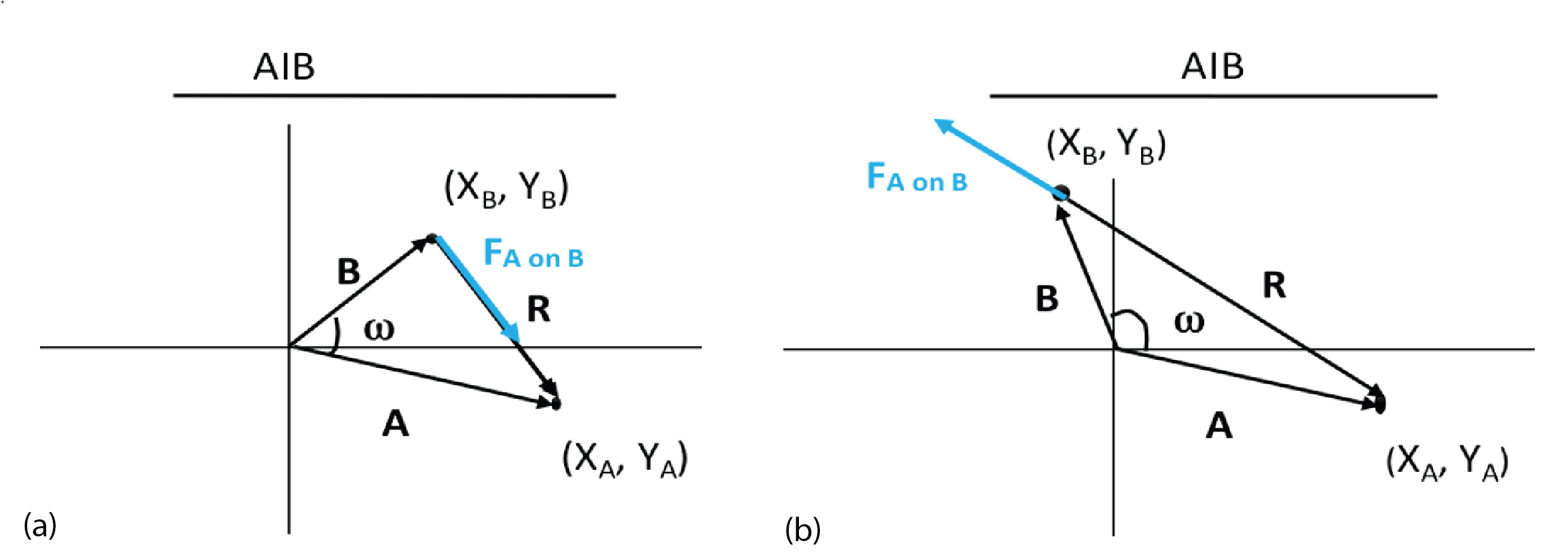} 
\vspace{5pt}
\caption{Vectors $\bm{A}$ and $\bm{B}$ represent agents in attribute space. $\bm{R}$ is the vector distance between agents and represents their difference in attributes.  The force $\bm{F}_{AonB}$ of agent $\bm{A}$ on $\bm{B}$ is shown in blue. The agents are allowed to communicate since $R<$AIB. a) The force is attractive since $\omega < 90\degree$; b) The force is repulsive since $\omega > 90\degree$.}
\label{fig1}
\end{figure}

\begin{figure}[h]
\hspace{2pt}
\includegraphics[scale=0.66]{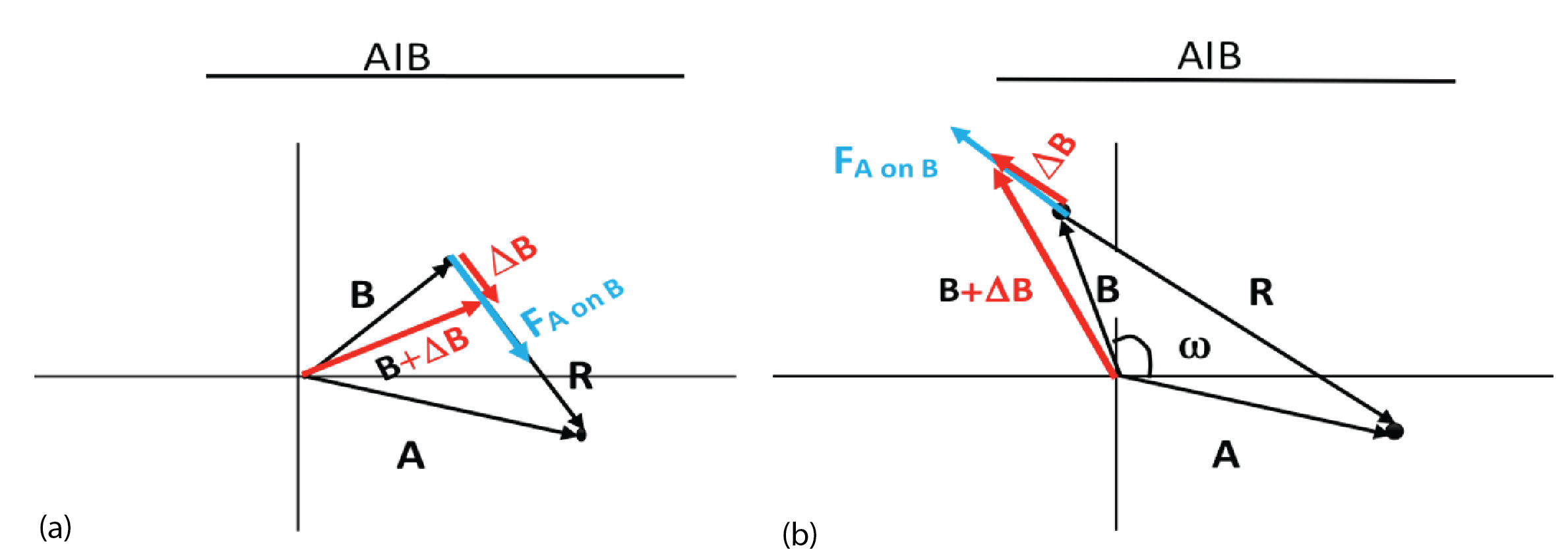}
\caption{The force $\bm{F}_{AonB}$ in blue changes vector $\bm{B}$ to $\bm{B}+\Delta \bm{B}$ in red.
a) attractive force which decreases the angle between $\bm{A}$ and $\bm{B}$; b) repulsive force increases the angle.}
\label{fig2}
\end{figure}


The forces between two agents affect the vectors representing the agents; the vectors of both agents are modified by their mutual influence. The force of agent A on agent B is a vector (vectors are in boldface type):
\begin{equation}
\textbf{F}_{AonB} = g m_{A} \frac{\textbf{A}\cdot\textbf{B}}{R^2} \frac{\textbf{R}}{R}  
\end{equation}
where $m_A$ is the active mass of agent A. The vector for agent A is $\textbf{A}=(x_A, y_A)$ and for B is $\textbf{B}=(x_B,y_B)$, and $\textbf{R}$ is 
\begin{equation}\space  \textbf{R} =\textbf{A}-\textbf{B},
\end{equation}
representing the vector separation between \textbf{A} and \textbf{B}. Note that if we are calculating the force of agent B on A, the direction of \textbf{R} is reversed but the magnitude is unchanged.  The magnitude of the separation R is
\begin{equation}
 R=\sqrt{g_x(x_A-x_B)^2+g_y(y_A-y_B)^2}   
\end{equation}
which we interpret as a numerical indication of the difference in the attributes for the two agents. If R is bigger than the AIB, the Attribute Influence Bound, then the agents are forbidden to interact and the force is set to zero.  The default maximum value of AIB is 280, which corresponds to the maximum distance between two agents in the visible universe.  By choosing expressions for $g_x$ and $g_y$ in Eq. 4, one can treat the $x$ and $y$ dimensions differently, which might be very useful in some applications. Cross terms in $x$ and $y$ are also possible.  To date we have assumed $g_x=g_y=1$.  In this case, the quantity $R$ represents the Euclidean distance in attribute space as indicated in Eq. 1.

The vector $\textbf{R}/R$ is a unit vector pointing from vector \textbf{B} to vector \textbf{A},  and $\textbf{A}\cdot\textbf{B}$ is the dot product of the vectors:
\begin{equation}
  \textbf{A}\cdot\textbf{B} = AB\cos\omega = x_A
  x_B + y_A y_B,  
\end{equation}  where $A$ is the magnitude of the vector $\bm{A}$: $A=(x_{A}^2 + y_{A}^2 )^{1/2}$, similarly for $\bm{B}$, and $\omega$ is the angle between the two vectors.

If $\omega$ is between -90 and +90 degrees, then the absolute value of $|\omega| < 90\degree $ and $\cos \omega$ is positive so the force is positive: the vectors have the same AO or Attribute Orientation and are considered as pointing in the same general direction. If $\cos \omega$ is negative, the vectors are seen as pointing in opposite directions and the force is negative. This is illustrated in Fig. 1. The AIB is also shown in Fig. 1, which is the maximum distance over which agents can influence each other.  

The parameter $g$ in Eq. 2 is an adjustable overall numerical factor that scales all forces equally, akin to the universal gravitational constant. The default value, $g=7.0 x 10^{-5},$ is chosen so typical forces result in a perceptible movement of the vectors in our space in a unit time interval $DT$. 

We choose the force to be proportional to $\textbf{A}\cdot\textbf{B}$ because this number is proportional to how parallel the vectors are:  $\textbf{A}\cdot\textbf{B}/B$ equals the component of $\textbf{A}$ that is parallel to $\textbf{B}$. The parallelism is interpreted as an indication of how similar or dissimilar the attributes of the two agents are, suggesting a general mechanism of cooperation/polarization  \cite{neirotti}\cite{brugnoli}\cite{newman}. 

The force is assumed to vary as $1/R^2$, in agreement with the force laws for gravity and electrostatics.  As discussed in Implementing the FVM in NetLogo, it is possible to adjust the exponent, although to date most of the simulations have assumed the exponent to be the default value of two. Increasing the exponent to three accentuates the role of short range forces, while reducing it to one, does the opposite. 

We assign a mass $m_A$ to each agent A which we call the active mass.  As shown in Eq. 2, the active mass $m_{A}$ scales the force or influence agent A has on another agent. The active mass might be considered analogous to an individual's charisma, the ability to influence others.

 \subsubsection*{Movement of the agents}
In order to compute the effect on agent B from the force $\textbf{F}_{AonB}$ we use the model of classical mechanics. We assume Newton's Law \textbf{F}=m\textbf{a} can be used to calculate the "acceleration" $\textbf{a}_B$ on B:

\begin{equation}
  \textbf{F}_{AonB} = M_B \textbf{a}_{B}  
\end{equation}
where $M_{B}$ is the passive or inertial mass of agent B, quantifying the "resistance" of B to influence or "acceleration".  
This constant acceleration $\textbf{a}_B$ acts for a time interval $DT$, during which it causes a displacement $\Delta  \textbf{B}$
in vector $\textbf{B}$

\begin{equation}
\Delta\textbf{B}=\frac{\textbf{F}_{AonB}}{M_B}(DT)^2
=g \frac{m_A}{M_B}\frac{\textbf{A}\cdot\textbf{B}}{R^2} \frac{\textbf{R}}{R} (DT)^2     
\end{equation}
and $DT$ is the duration of the time interval for this step in the simulation. This is a vector equation based on integrating Newton's Law with respect to time. In each iteration, changes to all agent vectors are calculated, then all vectors are updated simultaneously (Fig. 2), which for B takes the form
\begin{equation}
\textbf{B}\longrightarrow     \Delta\textbf{B} + \textbf{B}     .
\end{equation}

As shown in Eq. 7, the change in attributes $\Delta\textbf{B}$ of an agent B due to the force exerted by agent A goes directly with the active mass $m_A$ of agent $\textbf{A}$ and inversely with the passive or inertial mass $M_B$ of the agent being affected. The inertial mass or passive mass indicates the agent's resistance to influence, which has been described in model literature as anticonformity, contrarian or radical behavior, or stubbornness \cite{krueger}\cite{hegselmann2}. Agents who do not change their views significantly are agents with extremely high passive masses in our model, and have been termed contrarians. The bigger an agent's passive mass $M$, the less it will be affected by other agents.  This structure of $\Delta\textbf{B}$ in terms of active and passive masses is intended to reflect concepts of social influence.

In the FVM, note that as shown in Eq. 2 the force $\textbf{F}_{AonB}$ depends \textit{only on the active mass of} A, the charisma of A, and not on the charisma or active mass of B, and therefore does not equal $\textbf{F}_{BonA}$. This model is different from the ordinary equation for gravitational force between two masses, which depends on the symmetric product of the masses which implies that the force of B on A is the same as the force of A on B. As stated above, the asymmetry in the FVM is chosen in an effort to better capture the dynamics of social interactions.  

\subsubsection*{Adaptive time stepping}

In the computation of the change in coordinates for each agent for each iteration, the time interval $DT$ in Eq. 7 is initially set to 1, and the distance $\Delta\mathbf{B}$ to be moved by every agent is computed. The largest distance to be moved by any agent, the Peak Distance, is determined. Then the actual value of $DT$ to be used in the movement calculation Eq. 7 is scaled by the equation
\begin{equation}
 (DT)^2= CR/\text{Peak Distance}    
\end{equation}
where $CR$ is the Coalescence Radius.
This choice of $DT$ ensures that the maximum distance moved by any agent is equal to CR, the Coalescence Radius, which ensures smooth motion of agents in the simulation. Most agents will move a distance significantly less than CR. If forces are very small the Peak Distance would be less than the Coalescence Radius, then the time interval $DT$ would be increased from 1 to a value large enough to ensure that the maximum actual distance moved by any agent would equal the Coalescence Radius.  Adaptive time stepping reduces the number of iterations required and the time to perform a simulation.  

On the other hand, if the forces are very large, for example when agents are very near each other and about to merge into a group, then the value of $DT$ would be reduced below 1 to ensure that the maximum actual distance moved by any agent again equals $CR$, the Coalescence Radius.  

In many simulations, in the beginning agents are close to each other so the forces are typically large and the time interval $DT$ is often below 1 or between 1 and about 10. Later in the simulation, when groups are farther apart, the forces are smaller, and the time interval $DT$ becomes larger, 10 - 50, greatly reducing the number of iterations needed for a simulation while also ensuring an accurate simulation. In simulations with large AIB, the final phase of the simulation is typically dominated by small repulsive forces, with $DT$ having values of 100-170. In one iteration, a value of $DT$=100 gives the same movement in a simulation as $100^2=10,000$ iterations with $DT=1$.  Thus value of $DT$ is an indicator of the strength of the force.

The value of $DT$ for each iteration $i$, $DT_i$, is plotted during the simulation and the sum of the values of $DT_i$ for all iterations is displayed as the Elapsed Time for the simulation:  
\begin{equation}
 \text{Elapsed Time}=ET= \Sigma_i DT_i    
\end{equation}
where the summation is over the values of $DT$ for every iteration i to this point in the simulation.
The Elapsed Time $ET$ is a significant indicator of system evolution and reflects the forces and resulting movements of agents during the simulation. The ratio of Number of Iterations/ET=$NoIt/ET$ is an indication of the average strength of the forces in a simulation, and is a direct reflection of the role of adaptive time stepping.

\subsubsection*{Formation of groups}

If a positive or attractive force is causing the separation $R$ between two agents (group or individual) to decrease, then in each iteration the force between them will increase according to Eq. 2 and the distance between them will decrease by a coalescence radius $CR$ or less, according to Eq. 7.  Unless the simulation is stopped for some other reason, the distance $R$ between agents this close to each other can be expected to decrease to the $CR$ or less in several iterations.  If, after an iteration,  two agents are less than $CR$ apart, the agents will be merged into a group. The location of the group will be the location of the agent with the greater active mass. The active mass of the group will be the sum of the active masses of all members; the passive mass will be the sum of all the passive masses of the members.  This group will be treated like a single agent in subsequent iterations.  If other agents, individual or group, come within the $CR$ of this group, they will be joined to this group.  As explained in the Conceptual Background, the forces holding group members together are so dominant that groups are stable in the sense that no outside agent can remove an agent from a group.   

\subsubsection*{Role of the origin}

The origin plays a special role in the FVM. From the mathematical viewpoint, we have positive and negative attributes and the corresponding vectors with their base on the origin (0,0) can point in any direction. If two agents are in the same quadrant, they experience attractive forces.  If two agents are on opposite sides of the origin, for example in the first and third quadrant, they experience repulsive forces, so agents cannot cross the origin directly. Agents in adjacent quadrants may experience attractive or repulsive forces depending on the angle between them. 

To illustrate the role of the origin in simulations, consider the previous example of a universe consisting of agents that are described with attributes which range from extremely conservative to extremely liberal.  The coordinate system would indicate that conservative agents and liberal agents would tend to be on different sides of the origin. If a normal distribution of agents is centered on the origin, it will contain a sampling of agents up to a few standard deviations. Some agents will be conservatives and some will be liberals.  As the simulation proceeds, it will show the conservatives in one region forming groups, each representing the average opinion of its members and liberals will be forming groups, each representing the average opinion of its members.  This mean groups tend to form on opposite sides of the origin.  The conservatives and liberals tend not to mix in the simulation.  

On the other hand if the initial distribution of agents is some distance away from the origin, say in the region that corresponds to liberals, then this initial distribution contains primarily liberals.  As the simulation progresses, they will interact and move toward consensus among themselves. The effect of centering initial distributions away from the origin is discussed in more detail in Results.

As another example, consider a sample of a native population of some country, with their values and attributes, and an immigrant population, with significantly different values and attributes.  Imagine a coordinate system, in which the values of the native population tend to be in one region, on one side of the origin, and the values and attributes of the immigrant population tend to be on the other side of the origin.  A simulation with a random population, representing a mixture of natives and immigrants, would be centered on the origin. As the simulation evolves, groups would tend to form on either side of the origin, representing consensus among natives on one side, and consensus among immigrants on the other.

These examples illustrate that the outcome of the simulation depends very much on where the initial distribution of agents is placed.  An initial distribution centered about the origin will tend to show the formation of groups of roughly comparable size on either side of the origin.  An initial distribution of agents about a point some distance from the origin will generally show the formation of a dominant group representing consensus. The dependence of outcomes on the initial distribution presents both problems and opportunities to the modeler.  The challenge is determining the appropriate location of the origin in the attribute space for the system at hand.  The opportunity is that a broad variety of behaviors, from polarization to consensus, is possible, increasing the possibility of having a simulation that matches observed behavior. 

Also note that an agent at the origin of the coordinate system, (0,0), is characterized by a null vector, and so in our formalism does not exert a force on any other agent, nor do other agents exert a force on it.  On one hand one might view this agent as so completely average, that it is invisible. 

\section*{Implementing the Force Vector Model in NetLogo}

In this section, we discuss the operation of the FVM in NetLogo, including setting initial conditions for the agents, setting the parameters of the model, like g and $CR$, and the choice of default values. We also discuss the variety of conditions that can stop simulations and the quantities monitored during the simulation.  The simulations were run on an AMD Ryzen 9 3900X 3.8 GHz 12-Core Processor with 64 GB of memory running Linuxmint 20 (Ulyana) 64-bit system. We used NetLogo 6.1.1 for the simulation environment. Some of the simulations were verified on MacOS system running MacOS 10.13.06 with a 2.6 GHz Intel Core i7 processor. 

\subsubsection*{Setting the initial conditions for simulations }

Two key parameters that need to be set are N, the number of agents, and the AIB.  Typically N is 100-200, but may be more; larger numbers just require more time for the simulations. AIB can range from 0 to a default maximum of 280.  By resetting the slider in NetLogo, the maximum may be made as large as desired.  It is possible to expand the dimensions of the visible universe and then to be able to display the trajectories of all agents. The properties of a ghost agent, including location and mass, can be determined by inspecting it.

The other initial conditions specify the distribution of agents in attribute space and the masses of the agents.  Currently we have uniform random, normal, or exponential distributions of the initial locations of agents within the 200 x 200 visible space.  The parameters for the spatial distributions are set separately for the x and y dimensions.  For example, if a normal distribution is selected for x, then the x standard deviation and x mean must be indicated.  If a random distribution for y is selected, then agents will follow a uniform random distribution up to the edges of the entire visible universe $y=\pm 100$. Hence simulations with uniform random initial conditions will generally reflect the rectangular coordinate system.

The initial active and passive masses also can follow a uniform random, normal, or exponential distribution. It is necessary to set a value for the maximum mass that is allowed for all distributions. For the normal distribution, one must set the mean mass, and the standard deviation. One can also select to have active and passive masses for each agent that are equal.  In addition it is possible to manually create individual agents of any desired mass at any chosen location. By a separate setting in the dashboard it is possible to inject a defined number of agents all with the same mass, automatically placed with the same spatial distribution as specified for the other agents. It is also possible to create very regular and symmetric arrangements of agents by turning on the feature titled Geometry or to inject a single agent of any mass at a selected location .

If uniform random or normal spatial or mass distributions are specified, the random numbers required for a simulation are all generated from a single numerical seed which can be set in a window in the model interface.  Any non-zero integer whose absolute value is less than $2^{31}$ is suitable. If the value in the window is set to 0, then a new seed is automatically chosen for each simulation, and so a new random arrangement is generated.  If the value in the window is set to an integer (not to 0), then that selected number will be used for each simulation requiring a randomly generated initial arrangement.  This allows one to have exactly the same normal or random arrangement of agents for each simulation, while changing other parameters, such as the AIB.  The simulations with the same seed will be identical unless some parameter is changed.   

If the random generator sequence gives a mass that exceeds the maximum mass specified, then that mass is rejected and a new random mass is generated.  If that mass is below the maximum mass specified, then that mass is used. This process is followed until all masses are generated. 

A similar process is followed for the generation of the uniform random distribution or normal distribution of the initial x and y coordinates for an agent.  A coordinate outside of the visible universe is rejected and a new coordinate is generated.  A counter indicates the number of agents that were rejected because they initially located out of bounds. It is clear that the requirement of having the initial agents within the boundary of the visible universe, or having a mass that does not exceed a defined limit is not mathematically compatible with a truly random or normal arrangement. A truly normal distribution may have some initial agents placed outside the boundary of the visible universe, and so this will be allowed in the next version of the FVM.

\subsubsection*{Setting parameters}
The following parameters need to be set and may remain at the same value for most simulations:
\begin{enumerate}[nosep]
    \item AdaptiveDT-default is ON indicating adaptive time stepping ON.  If off, $DT$=1.
    \item DistanceExponent, default is 2.0 so forces go as $1/R^{2}$.
    \item CoalescenceRadius=CR, default is 0.2 .
    \item $g$, force scaling parameter, default is $7.0 x 10^{-5}$.
    \item Track, ON means agents leave a track during simulation, default is ON.
    \item Force Adjust, default is NORMAL, meaning forces are positive and negative as determined by the angle between agent vectors.  Other settings allow only positive forces(take the absolute value of force), only negative  forces (take minus the absolute value), no repulsive forces (set to zero), no attractive forces (set to zero), or inverting the sign of the forces.
      \item ShowForces-ON displays force vectors acting on agents during simulation; OFF no display. The length of the force vectors is scaled in each iteration.
\end{enumerate}

\subsubsection*{Termination of simulations}

A simulation may end by itself, for example, when all agents are separated by more than the AIB, or a simulation may be intentionally stopped by a variety of conditions, for example, when a certain number of agents remain. A window in the interface reports the specific reason the simulation ended. 

Simulations are run until: 
\begin{enumerate}[nosep]
\item Equilibrium is reached and there is no more movement of agents because there are no net forces.
\item All agents are separated by more than the AIB.
\item "tickstop" was set, indicating the maximum number of iterations requested was reached. Set to 0 to turn off. 
\item "AgentStop" was set, and the stop occurred when the indicated number of agents remained. Set to 0 to turn off.
\item "DTstop" was set, and the indicated maximum allowed value of DT was reached.  Set to 0 to turn off.
\medskip
\end{enumerate}

The simulations are designed to display the evolution of the system in time. As a consequence, one may want to explore the state of the system at different times and not just explore the end point. Generally it is possible to restart the same simulation by increasing the number in the appropriate window.

From observing many simulations with our Standard Conditions (defined below in Results), we see the values of DT vary from about 0.5 to 160, meaning that the corresponding forces vary by a factor of $(160/0.5)^{2}=102,000$. If DTstop is set, it means that a limit has been placed on how big DT can become and therefore how small the forces can become before they are not allowed. Setting a value in DTstop tends to end the simulation earlier. For example, one may want to put a upper bound on DT 
for a simulation of a real system that is subject to uncontrolled external influences that are not included in the simulation. There may be a point in the simulation when it is believed that these external forces have a comparable effect to the social interactions.  At this point, one may want to stop the simulation.  One way to do so would be to enter a value of DT as a limit.  
\subsubsection*{Quantities monitored during a simulation}

The Center Of Attributes $(X_C, Y_C)$ is continuously tracked as a small square during the simulation.  We define the Center of Attributes for all agents in direct analogy to the center of mass $(X_C, Y_C)$ for a flat object, modeled as an array of masses 
\begin{equation}
 X_C=\frac{1}{M}\sum_i m_i x_i \hspace{15pt} Y_C=\frac{1}{M}\sum_i m_i y_i 
\end{equation} where $M=\sum m_i$ is the total mass. In our case, $m_i$ is the mass of agent or group $i$ at location $(x_i, y_i)$, and $(X_C, Y_C)$ it can be interpreted as the Center of Attributes, an active mass weighted average of attributes, for all agents and groups. The mass serves as a proxy for the number of agents within a group.

Numerous quantities are calculated at each iteration and displayed during a simulation, including the number of all agents, number of groups, the mean distance between all agents  $ \frac{1}{N(N-1)}\sum_{i,j}^N |R_i - R_j|) $ where $R_i$ is the vector representing agent i, the mean distance from an agent to the closest other agent (mean distance to closest neighbor MDCN), the mean distance of all agents to the origin, a histogram of the angular distribution of agents, the annular distribution of agents (a histogram of agents as a function of their distance to the origin), a histogram of the mass distribution of agents, the maximum force on an agent, the maximum distance an agent would move if $DT=1$ (peak distance), the sum of all the attractive forces, and the sum of all repulsive forces, the ratio of the total repulsive force to the total of all forces times 100, the value of $DT$, the elapsed time (sum of all values of DT), and the moment of inertia. \medskip

 \section*{Results}

The evolution of the system is deterministic: the end point is determined by the exact placement and masses of agents in the initial distribution, by the choice of AIB and the other model parameters, including $CR$, $g$, and the distance exponent. In Results we first discuss Model Validation, treating NetLogo performance, the symmetries of the model, and the effects of variations in the key parameters of the FVM.

After Validation, we discuss numerous Simulations done with our Standard Conditions, meaning 100 agents with exactly the same normal distribution in x and y, with a mean of (0,0), and a standard deviation SD of 30 units in x and in y. The mass distribution is also normal with a mean of 60 and SD of 15, with active and passive masses equal, and we use the default values of g and CR ($CR=0.2$, $g= 7.0 x 10^{-5}$,  $1/r^{2}$). Adaptive time stepping is turned on for every simulation.

 \subsection*{Model Validation}

\subsubsection*{NetLogo Performance}
 NetLogo does all calculations with 64 bit precision, following IEEE 754 precision protocol for floating point numbers.  Results are given to 1 part in $10^{15}$. Multiple runs with identical input parameters have produced simulation results that are identical to 14 decimal places, indicating a stable computation. Simulations were done with different values of $g$, with a variety of values of $CR$ as well as many values of AIB, and no computational instabilities, such as those due to floating point errors were observed \cite{hegselmann2}. 

On the other hand, for some simulations done with the Geometry feature, with exact symmetric arrangements and very precise angles between agents, we occasionally noted potential computational issues, for example, the absence of exact repeatability, which we suspect may be due to floating point issues. These idealized situations have not been observed in simulations with Standard Conditions. 

A single NetLogo simulation with 100 agents, displaying the movements with tracks, and all other indicators, takes about 90 seconds. Using batch mode to execute simulations, with no visual display but recording all information, it took about 43 hours to do 120,000 simulations or about 1.3 seconds per simulation. 
 
\subsubsection*{Geometrical arrangements} Near the end of simulations with AIB $\geq 80$, there are generally two to four groups arranged approximately symmetrically around the origin.  It is observed that at this point in the simulations, repulsive forces begin to dominate.  To understand this behavior, we analyzed the forces present in numerous symmetric arrangements of agents. For example, consider a perfect square centered about the origin. The angle between agents on adjacent corners is 90\degree  so the corresponding force is zero, but the angle between agents on opposite corners is 180 \degree so the net force is a maximum repulsive force along the diagonal.  Thus the symmetrical repulsive forces result in an expanding square, with the agents going into the corners of the universe. For an equilateral triangle centered on the origin, the angles between vertices are all 120 \degree so the forces are all repulsive.  Since they are symmetric the net forces result in the uniform expansion of the triangle. FVM simulations of these geometrical arrangements agree with the calculations, and explain the behaviors often observed during simulations. These geometrical arrangements and other symmetrical arrangements can be modeled in the NetLogo FVM using the Geometry feature.

\subsubsection*{Rotational Symmetry}

 The FVM is based on a simple force equation with a dot product of vectors Eq. 2, which is a vector equation.  The coordinate system determines the components of the vectors representing the agents, but if the coordinate system were rotated, the forces between the agents, the angle between vectors, and the movement of agents as given in Eq. 2 and Eq. 7 would not change with this rotation. As a consequence the FVM has rotational symmetry. This means, for example, for the square arrangement of agents mentioned above, that the sides of the square centered on the origin can be at any angle with respect to the axes and the same dynamics of expansion of the square will occur relative to its corners. 

 To verify the presence of rotational symmetry with our Standard Conditions, we did a scatter plot of the final locations of the 2660 agents for 1000 simulations with AIB=100 (Fig. 3a). The size of each point corresponds to the mass of the agent. 
 
To determine if this distribution is independent of angle we measured the number of agents landing in 12 sectors each 30\degree wide and plotted the results in Fig. 3b. The data were normalized so that an ordinate value of 1 means that the sector received the number of agents predicted from a perfectly random distribution over 360\degree.  These figures clearly verifies rotational symmetry as expected from the mathematical properties of the model.

From the data, it is clear that the specific details of a particular initial normal arrangement of agents in a given simulation do indeed favor a particular outcome in which the final agents are located at specific angles relative to the vertical axis.  But when the results for many trials with different seeds are averaged, no angular bias is present, validating this feature of the model.   

Fig. 4a shows the distribution of distance from the origin for these same agents remaining at the end of simulations with AIB=100. The distance is approximately a normal distribution about a mean of 52.6, which is about half the AIB of 100, and a SD of 8.2. Fig. 4b shows that the mean distance to the closest neighbor MDCN at the end of the simulation is about equal to the AIB of 100 in about half the simulations. Thus as we have pointed out, at the end of the simulations, we typically have 2 or 3 agents separated by AIB from each other, on opposite sides of the origin. Further there is an approximately normal distribution of the masses of the final agents with a standard deviation of about 500.      

\begin{figure}[h]
    \includegraphics[scale=0.65
    ]{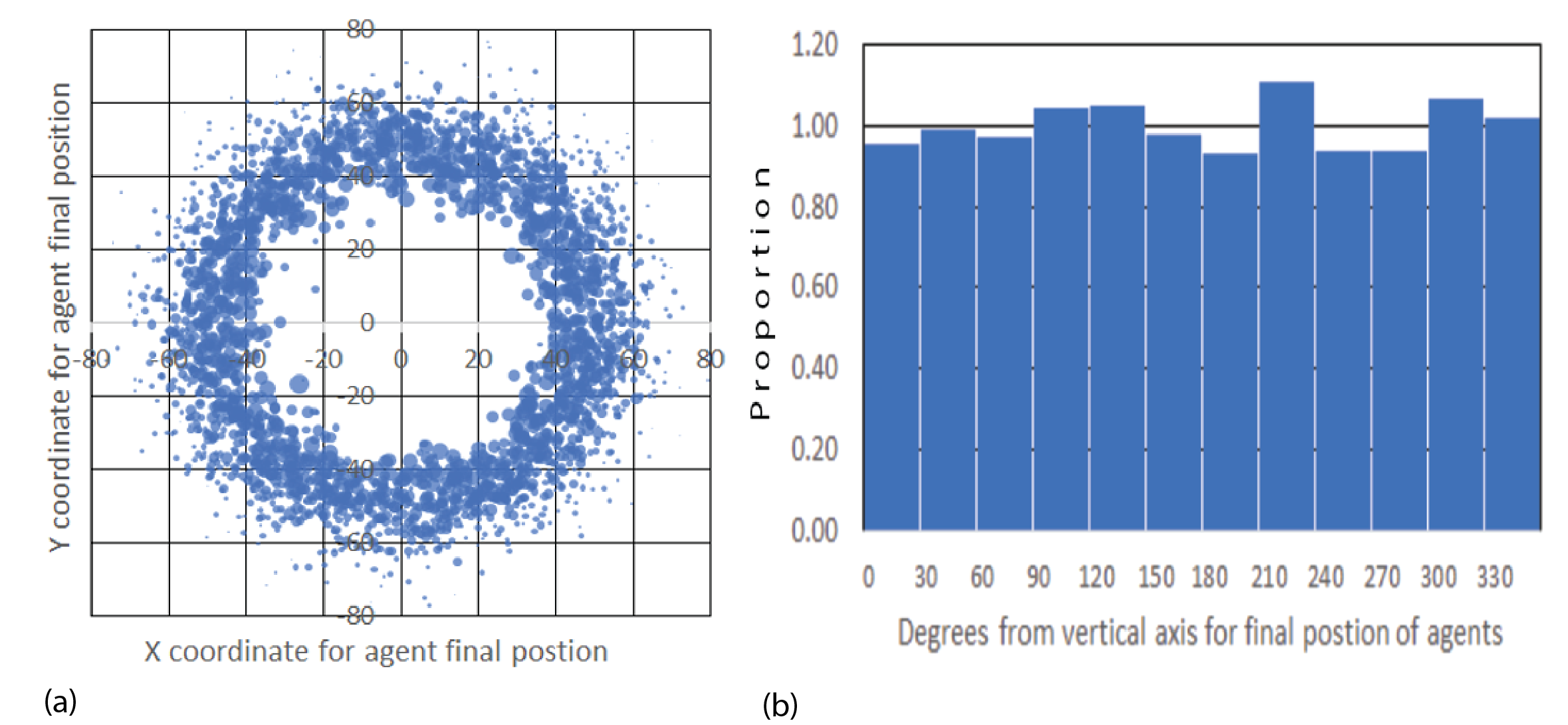}
    \vspace{5pt}
 \caption{Distribution of Agents at the end of 1000 simulations for AIB=100, Standard Conditions: a) Scatter plot of final location. The size of each point correlates with the agent mass; b) Angular distribution in sectors of 30 degrees. 1 means random distribution. The mean for all 12 sectors is 1.00 (random) and the SD 0.06.}
     \label{fig3}
 \end{figure}

\begin{figure}[h]
    \includegraphics[scale=0.64]{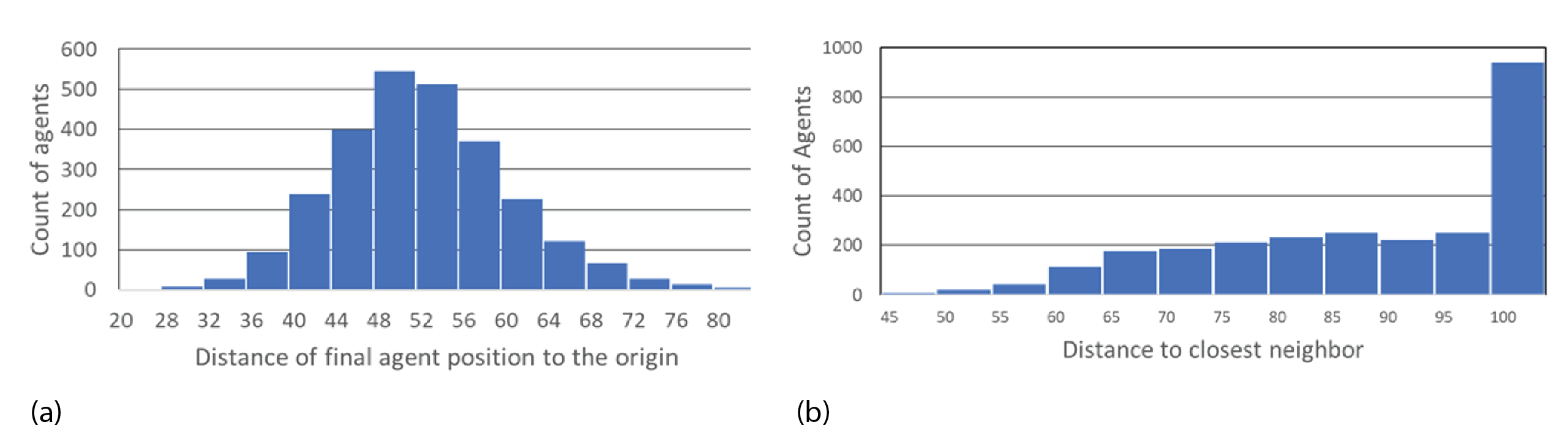}
    \vspace{5pt}
    \caption{Distribution of agents at the end of 1000 simulations for AIB=100, Standard conditions. a) Distance from the origin. The mean distance is 52.6, about half of AIB=100, and the SD 8.2; and b) MDCN=mean distance to closest neighbor. Median MDCN = 98.}
    \label{fig4}
\end{figure}

\subsubsection*{Center of Attributes}

For the 1000 runs depicted in Fig. 2, with Standard Conditions the average of the Center of Attributes $<(X_C,Y_C)>$ was initially about (0,0) with a SD of about 4 and at the end of the simulation averaged $\approx (0,0)$ with SD of about 5.  Thus in a given simulation the initial location for the Center of Mass could easily be (-2,4), for example, although the target was (0,0). During a simulation with Standard Conditions the Center of Attributes does not change significantly, typically by less than 6 units out of 200 in the visible universe.

Fortunato and his collaborators \cite{FORTUNATO_2005} described a similar conservation of the average opinion for their expanded version of the bounded confidence HK model of opinion dynamics\cite{hegselmann}. However, in their model, unlike in ours, this symmetry implied perfectly symmetrical arrangements of the agents at the end of the simulation, for example, consensus at the center or two perfectly symmetric clusters on either side of the center.

\subsubsection*{Effect of variation of parameters}
\begin{enumerate}
 
\item  Coalescence Radius: 

We have determined the effect of a variation in $CR$ by comparing simulations (Fig. 5) with different values of $CR$. Fig. 5a shows an overlay of simulations with $CR$= $0.1$, $0.2$, $0.3$, values bracketing the default value of 0.2. Fig. 5b shows an overlay with $CR$ = $0.2$, $0.5$, $1.5$. The tracks in the image show the movement of agents during the entire simulation, with the location of the final groups at the end of the simulation shown as triangles.  In an overlay, each simulation is shown in one color, and then different colored simulations are overlayed and oriented at the pixel level with respect to the fiducials of attribute space. For a variation from $0.1$ to $0.3$, Fig. 5a, there is very minor variation in the simulations, and the simulations are stable with respect to the small changes in $CR$. Simulations with other seeds show no detectable differences. For the greater values of $CR$, Fig. 5b, there is a small but clearly noticeable effect on the simulations and treatment of groups. 

Because of adaptive time stepping the value of $CR(ET)^2$ for a simulation remains nearly constant for small changes in $CR$. This follows since in adaptive time stepping $DT$ is adjusted to obtain a movement of $CR$. If $CR$ is reduced, more time steps will be required. For the values of $CR$ bracketing the default value of $0.2$, $CR(ET)^2$ equals $2.0\times 10^8$, with a variation of less than $0.8\%$. For the larger values of $CR$, the variation was up to $16\%$. Larger values affect the formation of groups and the evolution of the simulation. 

\begin{figure}[ht]
   
   \hspace{15pt}
    \includegraphics[scale=0.68]{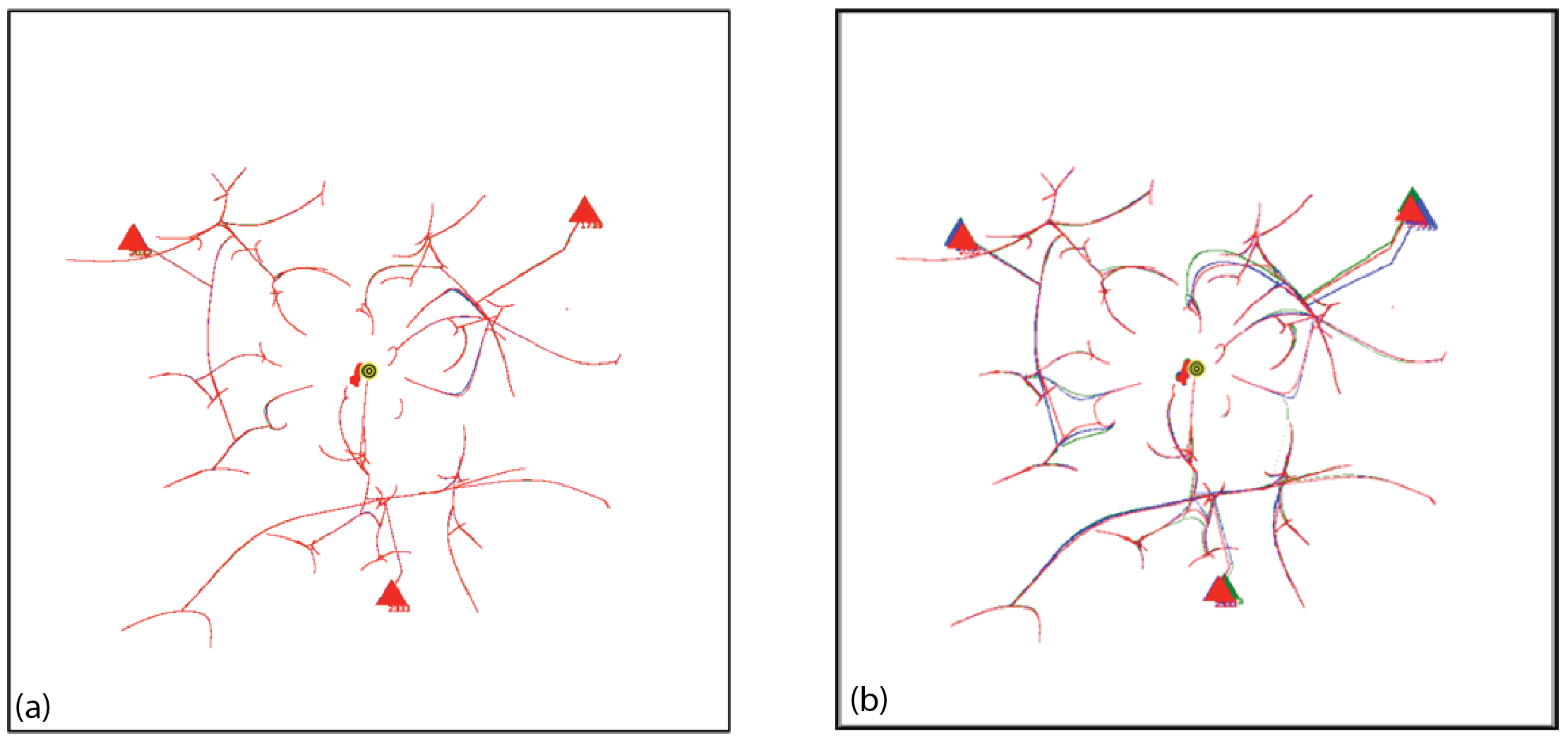}
    \caption{Overlay of multiple simulations to show the effect of variation in the Coalescence Radius $CR$, AIB=120, Standard Conditions, seed=61. A dark grey bullseye indicates the origin (0,0) in the center of the figure and the trail of a small red spot indicates the location of Center of Attributes during the simulation. a) Simulations with CR= 0.1(blue), 0.2(red/default), and 0.3(green). The red covers the blue and green tracks indicating they are almost identical; b)Simulations with $CR= 0.2$(red/default), $0.7$(blue), and $1.5$(green). Some differences are visible.}
    \label{fig5}
\end{figure}

\item  Force Dependence on Distance: 

We have done preliminary experiments with force laws varying as $1/r^{0.5}$, $1/r^{2}$ (default), $1/r^{5}$. Other than the change in the exponent of R, the force law is the same for each simulation.  Fig. 6 shows the agent tracks from each simulation; the colors are arbitrarily assigned to distinguish the initial agents. In Fig. 6c, there is an overlay of all tracks in one image. The AIB is 100, the seed 91.  Although the agent tracks are different, in all three cases the final separation between agents is 100, and the Center of Attributes changes by less than 5. Other seeds give different but comparable results. 

For the simulation with $1/r^{0.5}$ shown in Fig.6a,  the number of iterations $NoIt$ is 1411,  elapsed time $ET$ is 503, the ratio $NoIt/ET = 2.8$, which serves as an indication of the average strength of the forces, the larger the ratio the larger the average force. During the simulation the maximum $DT$ is only 0.5 and virtually all agents are moving together to the poles, like particles in the field of a magnet. Strong long range forces dominate the simulation which evolves comparatively quickly, with an elapsed time about 25,000 times shorter than for the $1/r^{5}$ simulation.

For $1/r^2$ shown in Fig. 6b, the number of iterations is $NoIt= 2277$, the $ET$ is 23,360, the ratio $NoIt/ET = 0.1$, indicating average weaker forces, $DT$ is generally less than 100. The repulsive force dominates at the end. The simulation shows some balance between the dominance of short range forces and long range forces during the simulation.  This is one reason why we use the $1/r^2$ force law as our default.

For $1/r^5$ in Fig. 6c, the number of iterations is $NoIt=3740$ iterations, $ET$ is $1.28 x 10^7$, 550 times greater than the $ET$ for $1/r^{2}$. $DT$ is 92000 at end, almost 1000 times bigger than for $1/r^{2}$, so the final forces are $10^6$ times weaker. Ratio $NoIt/ET= 0.0003$, indicating extremely weak average forces. The simulation starts with total local control and only the few very nearest agents merge.  Eventually, with very large values of $DT$, the very small forces for larger separations cause groups to coalesce and move inward.  The repulsive force drives separation to AIB at the very end.  

The end points of the three simulations are similar, two groups on opposite sides of the origin separated by AIB, yet the evolutions of the simulations are very different as Fig. 6d shows, presenting the overlay of the tracks for the three simulations. For example, if one stops the simulation when there are a certain number of agents, the arrangements will be significantly different for the three exponents.  The exponent effectively determines the relative weight given to long and short range interactions. The higher exponents significantly increase the effect of short range interactions over long range interactions causing the shortest range interactions to dominate. The higher exponents reduce the strength of longer range interactions. For a separation of 100, the forces for the $r^{5}$ exponent are $10^6$ times weaker than for the $r^2$ exponent, hence the dramatically longer elapsed time.

Why do the simulations with Standard Conditions end with agents in similar location even though the earlier steps are very different? Essentially for the same reason the simulations with $1/r^2$ often end with two agents on approximately opposite sides of the origin, or three agents in a triangle about the origin. The force law for all simulations is the same as in the standard FVM except the exponent of R is different, so the symmetries are the same, and the arrangement of agents in final states is a consequence of the overall symmetries: 1. The model is a vector model that shows no inherent preference for direction when the initial conditions are symmetric about the origin;  2. This means the final arrangement will also tend to be symmetric about the origin, for example, with two groups left, they will be approximately on opposite sides of the origin, with three groups, they will be approximately in an equilateral triangle, both arrangements in which repulsive forces dominate, and cause the separation to increase until the AIB is reached; 3) Adaptive time stepping compensates for lower forces, just increasing the elapsed time.

\begin{figure}
   \hspace{10pt}
    \includegraphics[scale=0.73]{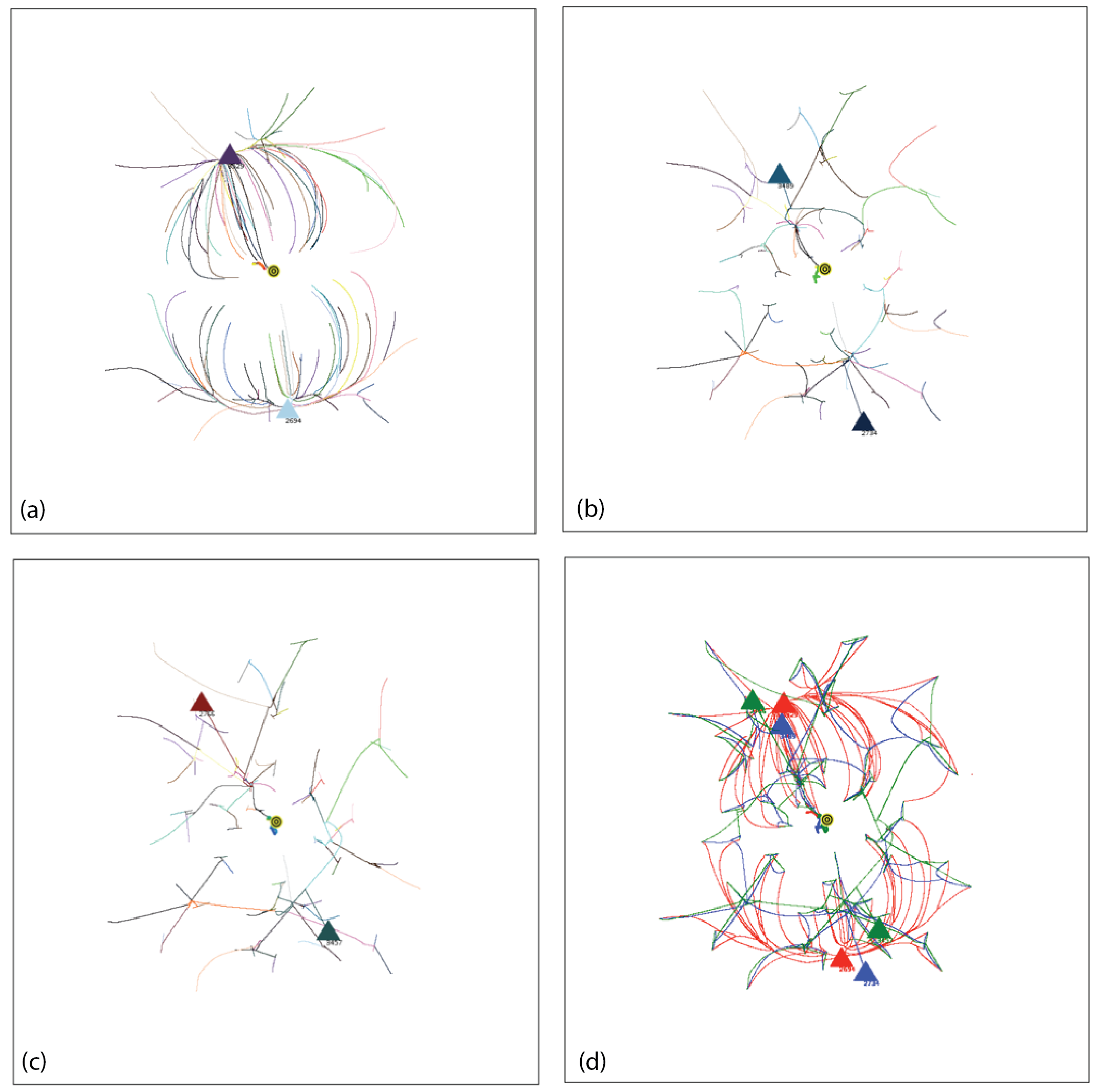}
    \caption{Simulations for the same FVM force vector law but with three values of the force exponent, with AIB=100, seed 91. (a) $1/r^{0.5}$. ET$=503$, $ DT \leq0.5$. (b) $1/r^{2}$. ET$=23,360$, $DT \leq 100$. (c) $1/r^{5}$. ET=$1.3\times!0^{7}$, $DT \leq 1.28\times 10^{7}$. (d) Overlay of tracks from all three simulations with red $r^{0.5}$, blue $r^{2}$, green $r^{5}$. The colors of the tracks in a,b and c correspond to the color assigned to each initial agent.}
    \label{fig6}
\end{figure}

\item Effect of variation in g: 

All forces on agents and all movements of agents are scaled by the strength parameter $g$, as shown in Eqs. 2 and 7.  We have done simulations with adaptive time stepping in which $g$ was about 1/3 of the default, equal to the default, and 4 times the default, and equal to 1.0, which is about $10^{4}$ times the default, as shown in Fig. 7.  The agent tracks in all simulations are absolutely identical, but the elapsed times are very different, ranging from about 57,346 for $g=2 x 10^{-5}$ to 256 for g=1.  Thus the agent trajectories and the essential results of the simulations are totally independent of $g$.  Eqs. 7 and 9 show that adaptive time stepping ensures $DT^2$ is chosen so all distances moved in an iteration are CR or less, no matter what the force, and with Eq. 10 for the elapsed time $ET$ it follows that for a given CR the quantity $g(ET)^2$ is a constant for all values of g. For each of the four cases considered $g(ET)^2$ is equal to $6.577255003 \times 10^4$ to one part in $10^{10}$ and the number of iterations is exactly 2580.  In each iteration the movement of the agents is identical for all g values.  The results of this experiment show that the adaptive time stepping is stable and working as designed. It also emphasizes that the outcome of a simulation does not depend on the choice of g since adaptive time stepping compensates for enormous differences in the forces between agents by increasing the elapsed time.  Changes in g change the rate at which the elapsed time passes during the simulation, but all agent tracks are unchanged.  If adaptive time stepping is turned off, then changing g will indeed change how far an agent moves in each iteration and will yield a different simulation. For larger values of g it will not produce a simulation that meets our objectives of continuity or group formation.

\begin{figure}
    \hspace{100pt}
    \vspace{10pt}
    \includegraphics[scale=0.36
    ]{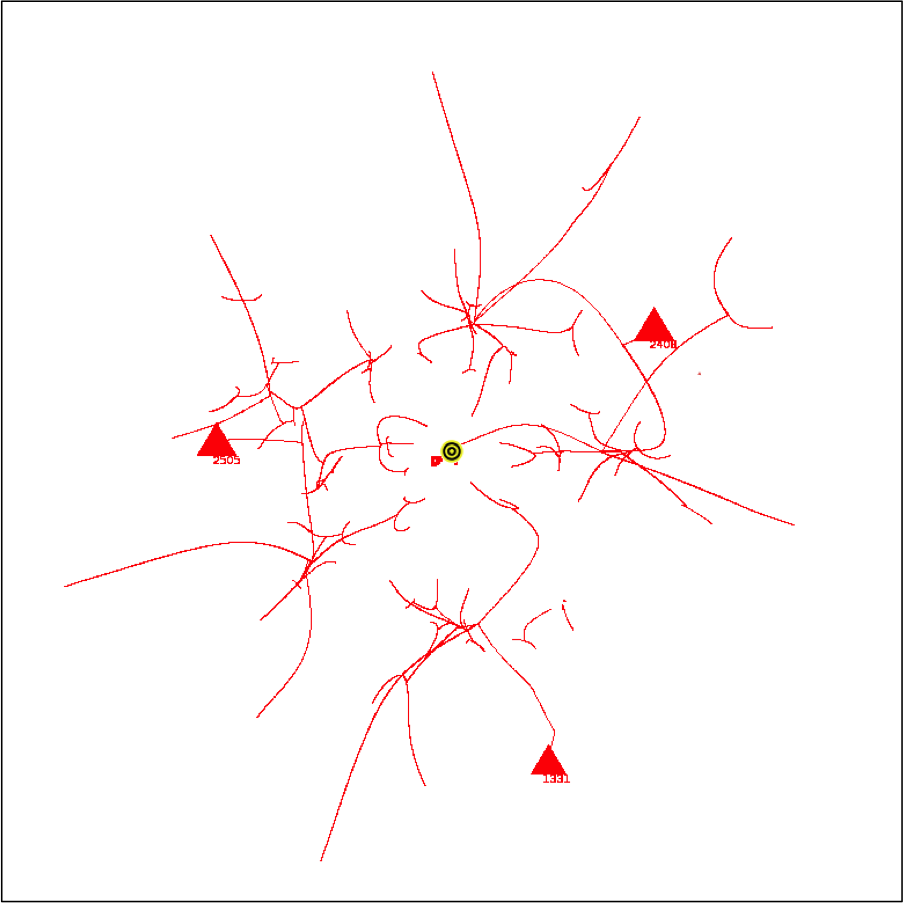}
    \caption{Overlay of simulations in red, blue, green, and brown, with four different values of g, all have $AIB=100$, seed$=71$ and 2580 iterations. The final red tracks cover the blue, green and brown tracks perfectly: 1) red $g=2.0\times 10^{-5}$, $ET= 57346$; 2) blue $g=7.0\times 10^{-5}$ (default), $ET= 30653$; 3) green $g=3.0 \times 10^{-4}$, $ET=14806$; 4) brown $g=1.0$, $ET=256.46$ }
    \label{fig7}
\end{figure} 

\end{enumerate}

\subsection*{Simulations with Standard Initial Conditions}

Because of the great variety of possible initial distributions of agent locations and masses, we use Standard Conditions as a baseline for exploring the properties of the FVM. Multiple runs are generated using different seeds for the statistical distributions so each run is unique.

\subsubsection*{General observations}
  
 Perhaps the most notable result is that for long timescales similar agents have coalesced into groups of generally comparable size and the groups tend to stabilize in distinctly different regions of the attribute space. The Mean Distance to the Closest Neighboring agent MDCN at the end of the simulation is approximately equal to the AIB, the Attribute Influence Bound. The greater the distance AIB over which agents can communicate, the greater the separation between agents at the end of a simulation. For large AIB, 2 or 3 groups of roughly comparable size remain at the end of a simulation and strong polarization is evident.  

Consider a typical simulation using Standard Conditions as shown in Figs. 8-11.  On the left, AIB=30, on the right AIB= 180. (For an animation of these simulations go to \url{https://youtu.be/WzblcRCjiK8} and \url{https://youtu.be/BArqDp8-JAQ} for AIB=30, 180 respectively.). At the beginning of the simulation, agents closest to each other tend to move towards each other because of the inverse square force law: forces are strongest between agents which are closest.  Note that an agent at an angle $\theta$ experiences attractive forces between all agents at an angle $\phi$ that is within plus or minus $90\degree$ of $\theta$:  $\theta-90\degree<\phi<\theta + 90\degree$. These initial attractive forces produce groups in representative niches in attribute space. After 10\% of the Elapsed Time to completion of the simulation, the arrangements look like Fig. 8a for AIB=30, with 53 agents remaining, and Fig. 8b for AIB=180, with only 10 groups remaining. The role of AIB is clearly illustrated here: with the small AIB, agents and groups can only interact with those that are relatively close so more groups remain at this stage, conversely with the large AIB, forces are present between distant groups that may be in different quadrants, and interactions, including those with repulsive forces can take place.

In the legends, we give some key statistics to characterize the simulation to this point in time and the resulting arrangement of agents, including the number of agents and groups, the MDCN. the strength of the repulsive force as a percent of the total force, and range of DT, which is an indicator of the overall strength of the forces: the bigger DT, the weaker the forces.

After 50\% completion, the simulations are shown in Fig. 9. For AIB=30, 20 agents remain, spread over much of attribute space. There are two outliers on the far left, a group and an individual and two groups on the far right that all remain in the same locations for the rest of the entire simulation since they are more distant than AIB=30.  Most of the consolidation is in the more densely populated central region. 

For AIB=180, at 50\% completion only the final three groups remain.  At this point, the attributes of each of the groups represents the average opinion of all of its members.  As the simulation proceeds, these groups will continue to move away from each other under repulsive forces (Figs. 10b,11b).  This is a typical ending for many large AIB simulations, where two or three groups separate symmetrically until they are AIB apart, and can no longer communicate.  In the last phase of separation between these groups, the groups become polarized, meaning that the attributes of the group become more extreme than the initial average of the attributes of its members. Notice the Elapsed Time for this large AIB simulation is about 6 times as great as that for the AIB=30 simulation, and that well over half of the elapsed time occurs during the separation of groups under repulsive forces, which are weak because of the large distance separating the groups. As this comparison illustrates, the Elapsed Time is approximately directly proportional to the AIB.

For AIB=30, after about three quarters of the Elapsed Time (Fig. 10a), the 16 agents that were in the central region have joined to form 12 groups. At 100\% completion (Fig. 11a), these 8 have joined to form 4 groups, spaced somewhat uniformly around the origin. The percentage of the total mass for the four groups near the center is, clockwise from the vertical: 27.5\%, 35\%, 14.5\%, 17\%, which is roughly a normal distribution (see Fig. 15). The 4 outliers account for about 6\%. The MDCN is 38, about 1.25 AIB. Each of the 4 groups represents attributes that are the average of the attributes of all of its members, which came from the immediate region around each group.  Since the mean mass of an agent is 60, the mass of a group is a proxy for the number of members of the group.


\begin{figure}[h]
\hspace{15pt}
\vspace{5pt}
\includegraphics[scale=0.6]{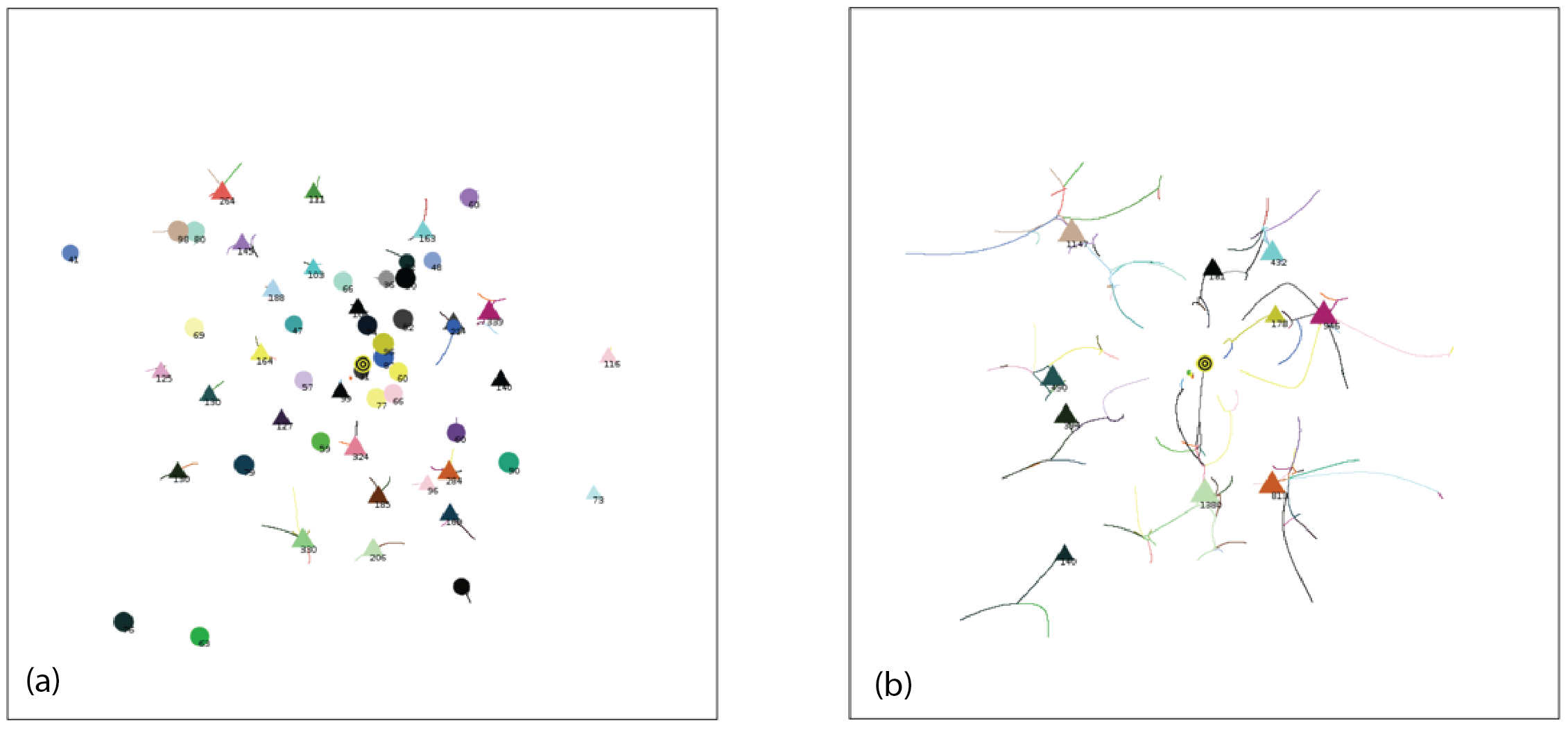}
\caption{Simulation at 10\% ET of completion. Groups=triangles, individual agents=circles.  Standard initial conditions. Seed=61. a) AIB=30, ET=680, 27 individuals(circles), 26 groups(triangles), MDCN=12, repulsive force$<0.5\%$, DT$<5$. b) AIB=180, ET=5490, 10 groups, MDCN=21, repulsive force$<20\%$, DT$<17$. For animation for 30 AIB: \url{https://youtu.be/WzblcRCjiK8}}

\label{fig8}
\end{figure}

\begin{figure}[h]
\hspace{15pt}
\vspace{5pt}
\includegraphics[scale=.6]{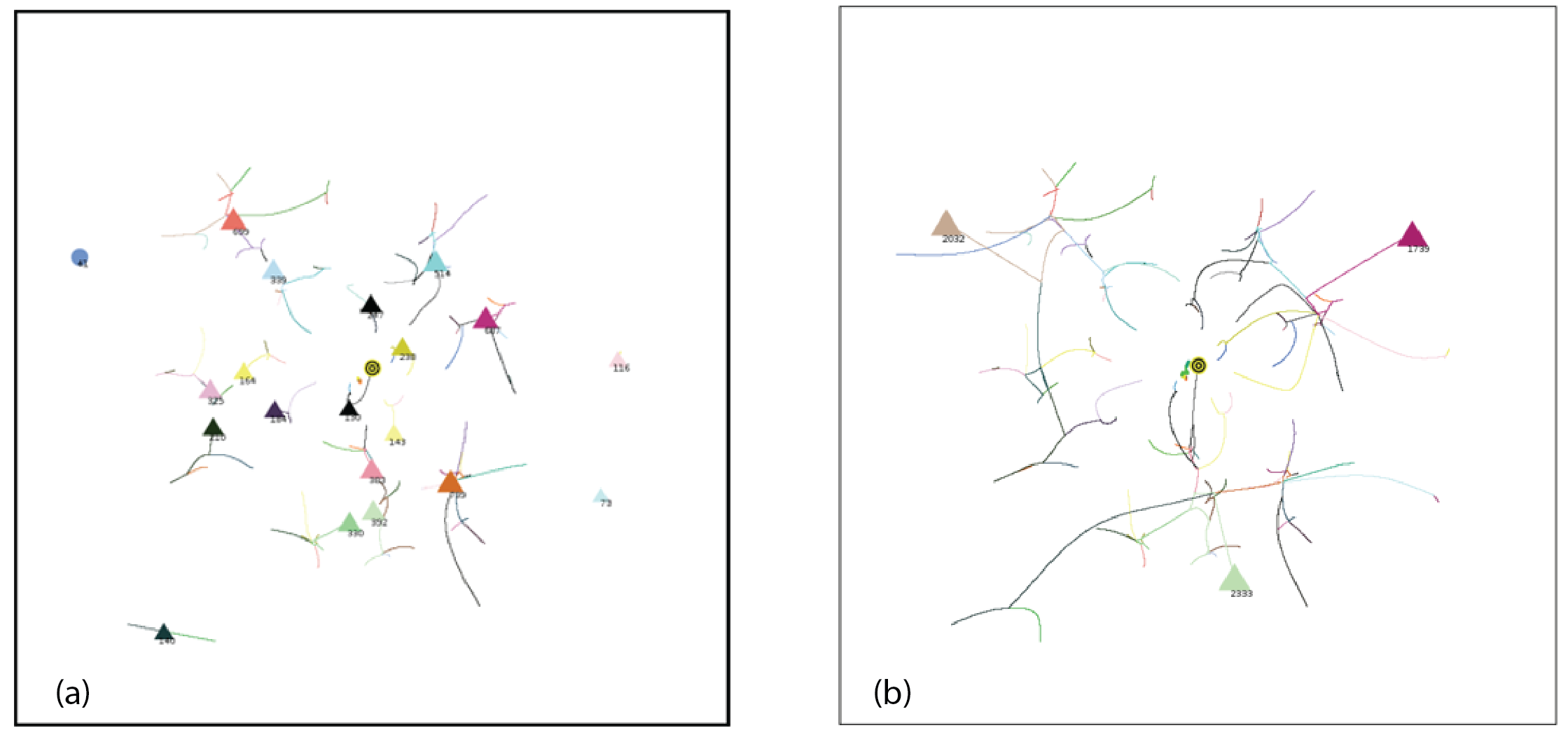} 
\caption{Simulation at 50\% ET of completion. a) AIB=30, ET=3410, 1 individual, 19 groups, MDCN=21, repulsive force$<0.8\%$, DT$<15$. b) AIB=180, ET=27041, 3 groups, MDCN=115, repulsive force$ \leq 100\%$, $DT\leq 110$.}
\label{Fig9}
\end{figure}


\begin{figure}
\hspace{15pt}
\vspace{5pt}
\includegraphics[scale=0.6]{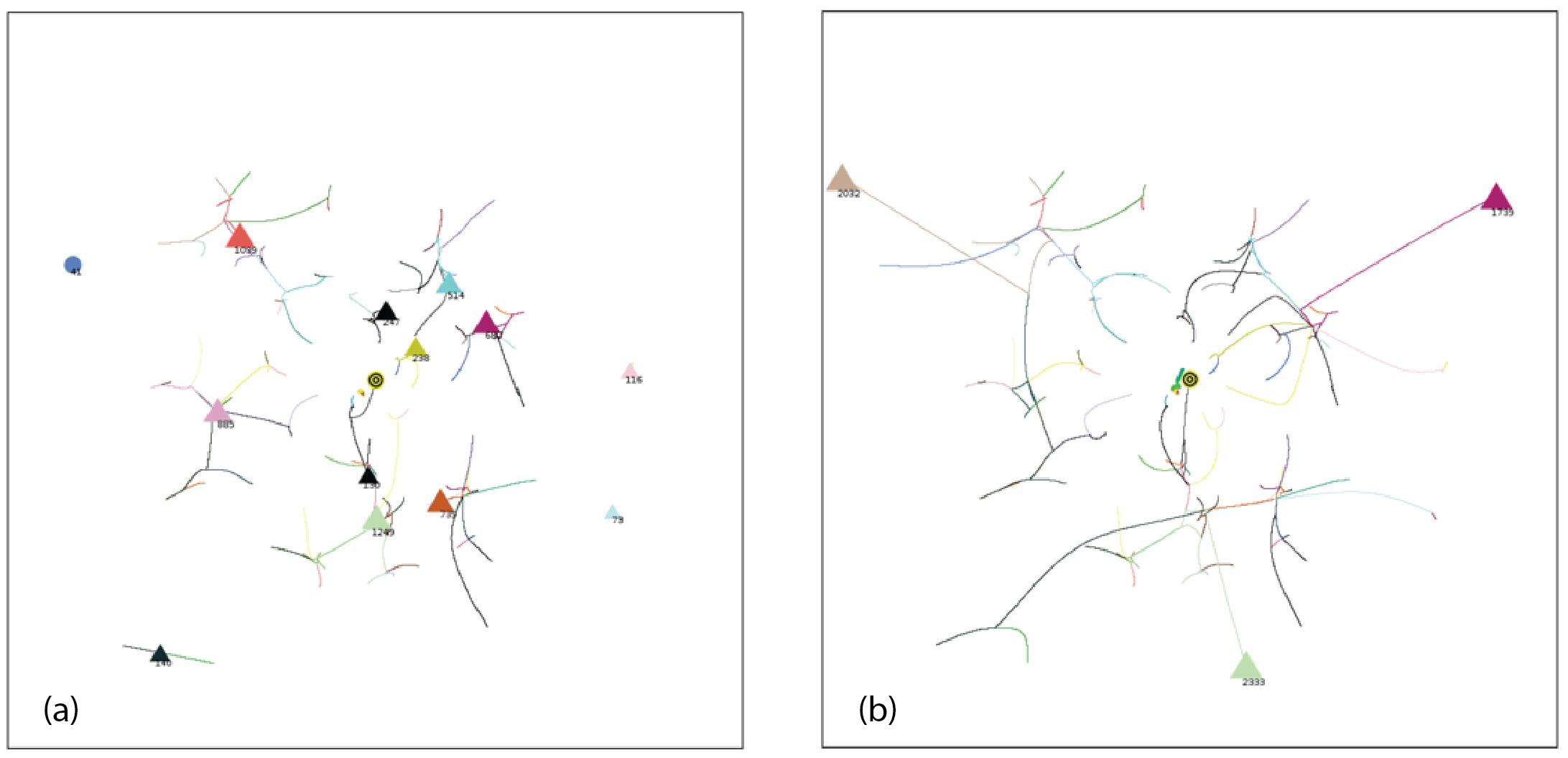}
\caption{Simulation at 75\% of completion. a) AIB=30, ET=5116, 12 groups, 1 individual, MDCN=28, repulsive force$ <0.8\%$, DT$<20$. b) AIB=180, ET=40,502, 3 groups, MDCN=154, repulsive force$ \leq 100\%$, $DT\leq 110$. For animation of this simulation with AIB=180 go to \url{https://youtu.be/BArqDp8-JAQ} }
\label{fig10}
\end{figure}


\begin{figure}[ht]
\hspace{15pt}
\vspace{5pt}
\includegraphics[scale=0.6 ]{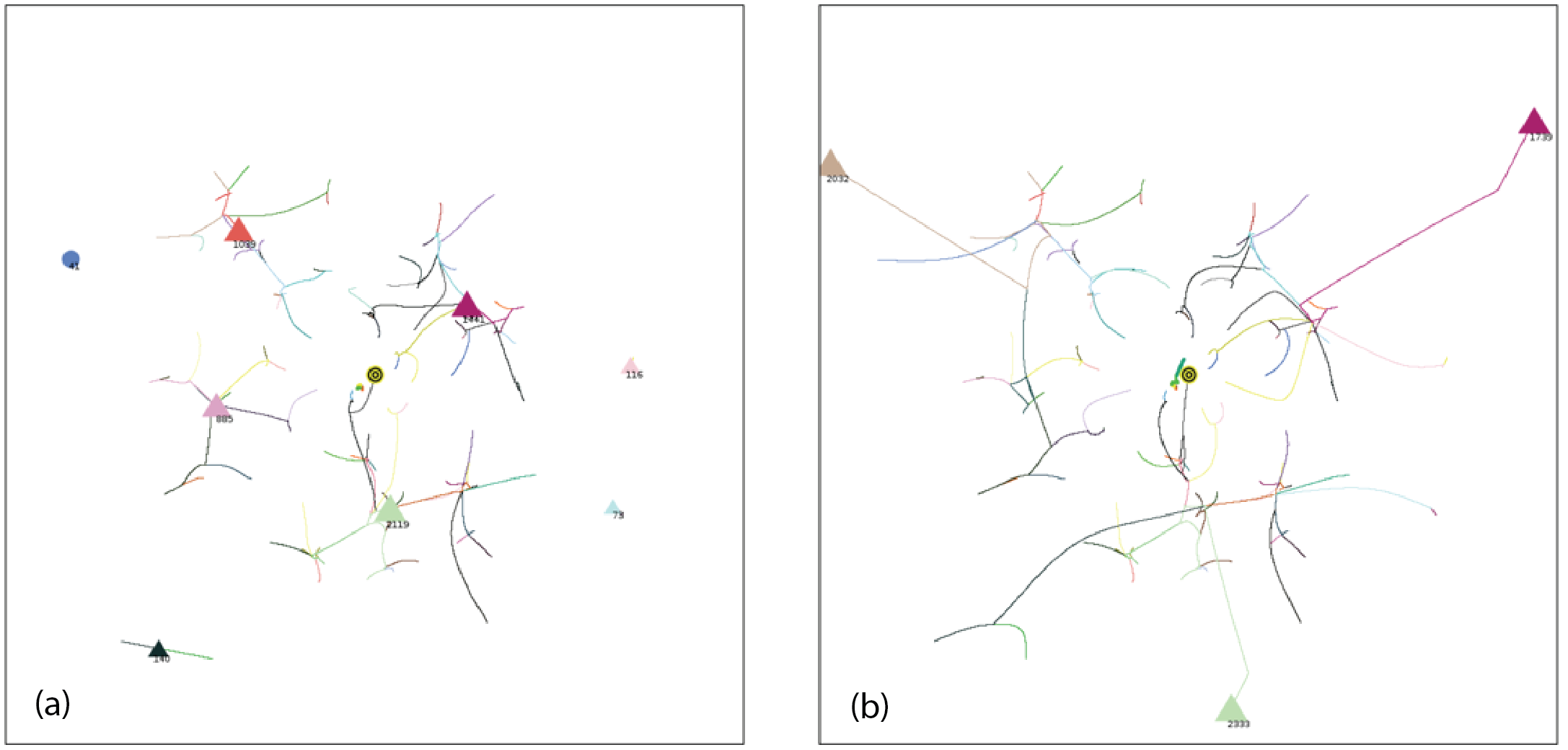} 
\caption{Simulation at 100\% of completion. a) AIB=30, ET=6821, 7 groups, 1 individual, MDCN=38, repulsive force$ <0.8\%$, DT$<20$. b) AIB=180, ET=54,030, 3 groups, MDCN=181, repulsive force$ =100\%$, $DT\leq 140$.}. 
\label{fig11}
\end{figure}

\begin{figure}[ht]
\hspace{13pt}
\vspace{5pt}
\includegraphics[scale=0.6]{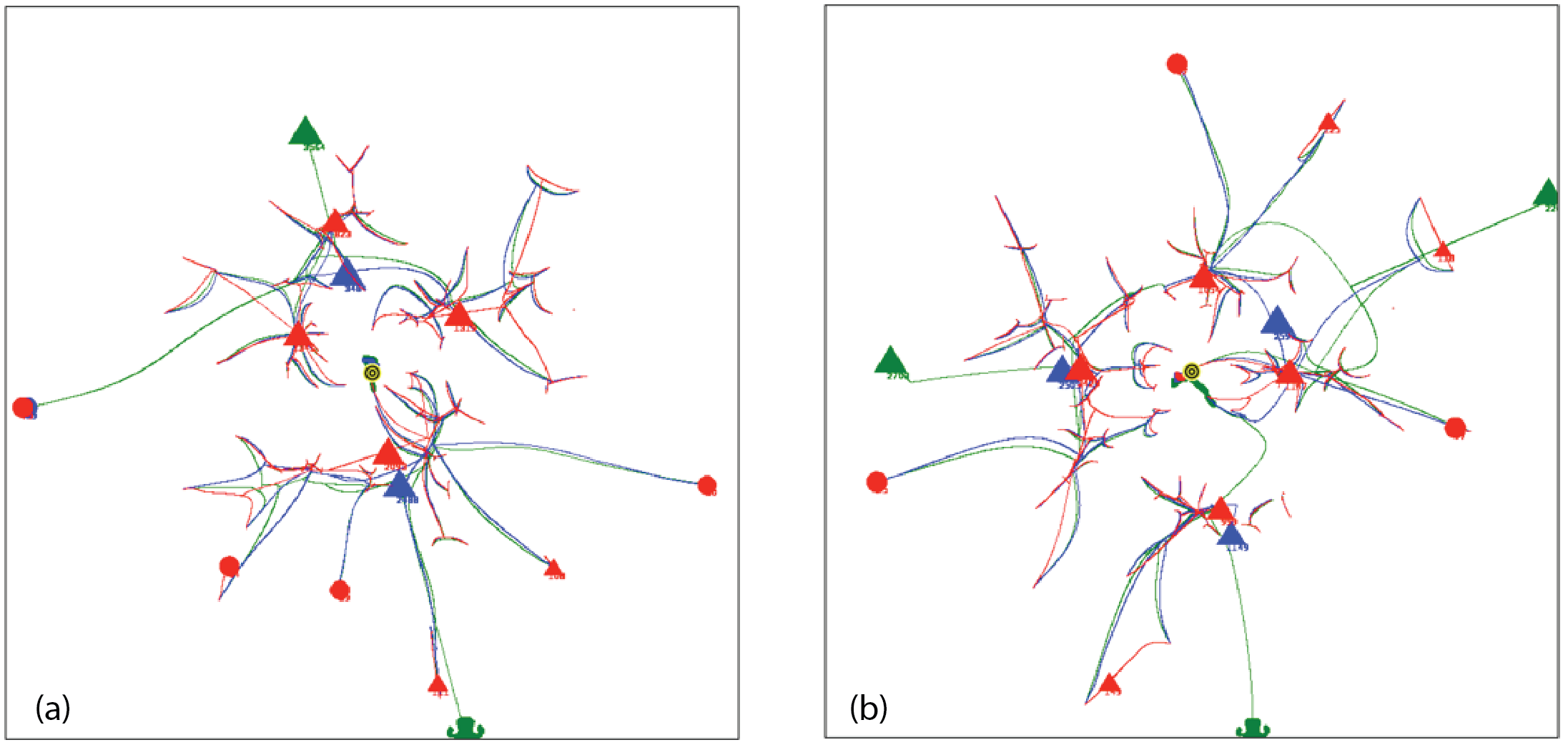} 

\caption{Overlay of simulations with AIB = 30(red), 60(blue), 180(green). a) Seed=$51$, Number of final agents= 10(red), 3(blue), 2(green). b) Seed= $71$. Number of final agents=10(red), 3(blue), 3(green).}
\label{fig12}
\end{figure}

Fig. 12 provides another illustration of the role of AIB in simulations with Standard Conditions. The figure shows an overlay of the final state for separate simulations with AIB of 30 (red), 60 (blue), 180 (green), for two different seeds (Fig. 12a and 12b). In an overlay, each simulation is shown in one color, and then different colored simulations are overlayed and oriented at the pixel level with respect to the fiducials of attribute space. For the small AIB, 10 red agents, widely spaced, are present at the end of the simulation; for an AIB of 60 these coalesce into 2 or 3 blue groups that are about 60 apart, spaced approximately symmetrically around the origin. For the AIB of 180, the blue groups separate from each other with a green trail under predominantly repulsive forces until they are about AIB=180 apart. The tracks of the agents during the simulations are similar, but not identical, for as long as each simulation lasts, but low AIB simulations end earlier since the Elapsed Time is approximately directly proportional to the AIB value. 

\subsubsection*{Number of agents remaining at the end of a simulation}

One of the important results of the model is the number of agents and groups that remain at the end of a simulation.  Fig. 13a shows the mean number of agents (individuals plus group) in the blue upper curve and groups in the lower red curve as a function of the Attribute Influence Bound AIB from 0 to 100 with our Standard Conditions.  The number of agents remaining decreases rapidly as AIB increases from 1 to about 60, where it is 2.68. From AIB 60 to 100 the number of agents decreases slightly to 2.25. Fig. 13b highlights the continued slight decrease to 2.06, with a standard deviation of 0.24, at AIB=280. For these plots 100 runs with different seeds were done for each value of AIB. 

The plot in Fig. 14 shows that the number of agents at the end of a simulation declines exponentially from 98.6 to 3.2 as the AIB increases from 1 to 52. For this range of AIB, the attractive forces tend to dominate, and repulsive forces typically account for from less than 1\% to at most 15\% of the total force in each iteration. Thus the coalescence of agents into groups is primarily due to attractive forces.

We have analyzed the angular distribution of the agents remaining at the end of the simulations used to obtain the data in Fig. 13 and we find regularities. If two agents remain, the average and median angle between them is $360/2=180\degree $ with a $SD=15 \degree$ and the average mass is $6000/2=3000$ with $SD=580$ (total mass is 6000=100 agents x 60/agent). If three agents remain, the average and median angle between adjacent agents is $360/3=120\degree$ with $SD=17.5 \degree$ and the average mass is $6000/3=2000$ with $SD=653$, with similar results for 4 and 5 agents.  An example of the frequency distribution of the angles between adjacent agents is shown in Fig. 15a for final states with 3 agents. The corresponding mass distribution is shown in Fig. 15b.

\begin{figure}
\vspace{1pt}
\includegraphics[scale=0.65]{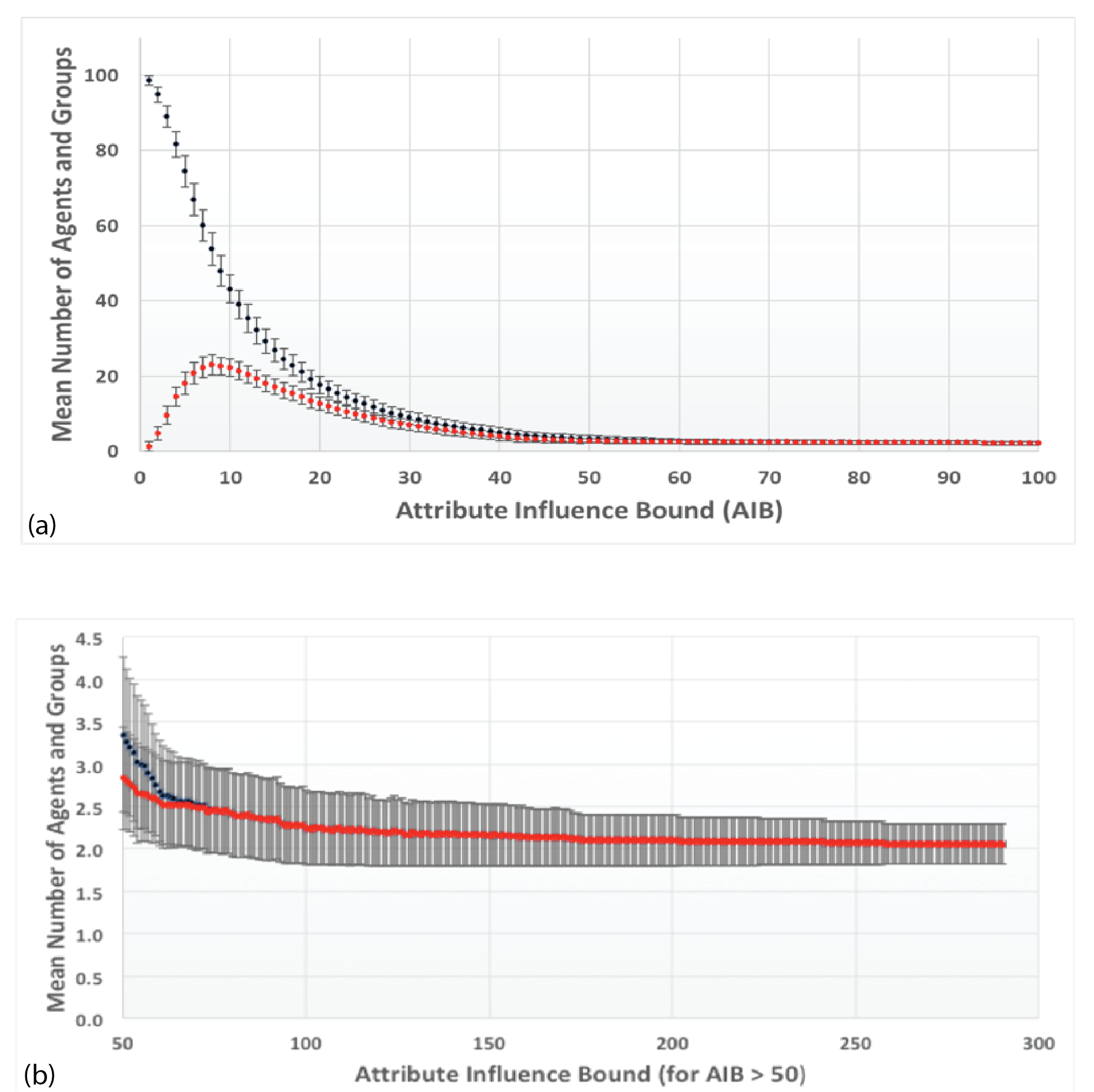} 
\caption{Mean number of agents (black) and groups (red) with standard deviations at the end of a simulation as a function of the Attribute Influence Bound (AIB). Standard Conditions, 100 runs/AIB value: a) AIB from 0 to 100; b) AIB from 50 to 280.}
\label{fig13}
\end{figure}

\begin{figure}
\hspace{10pt}
\includegraphics[scale=0.65]{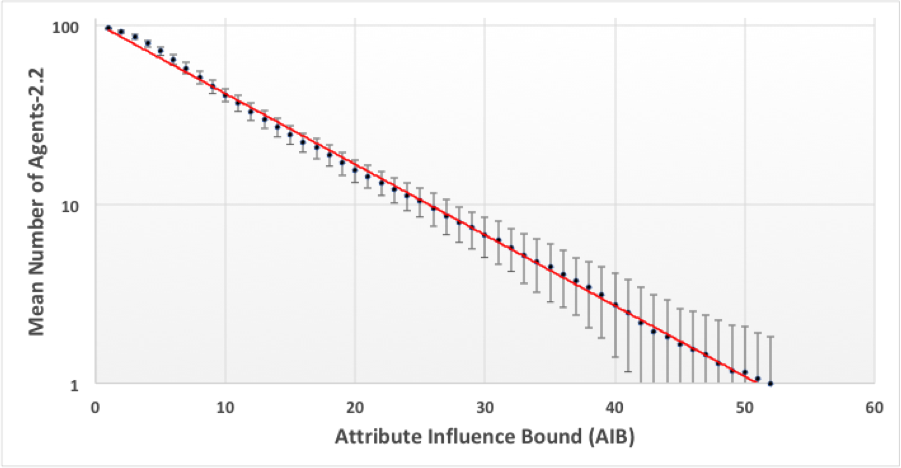} 
\caption{Ln(Mean Number of Agents - 2.2) versus AIB, Black points with standard deviation are data and Red is a linear fit. Standard Conditions.  Illustrates the exponential decrease in the number of final agents from 98.6 to 3.2 as AIB increases from 1 to 52.}
\label{fig14}
\end{figure}

\subsubsection*{Distance between agents at the end of a simulation}

We have also analyzed the relative locations of the agents remaining at the end of the simulations used to generate Fig. 13. Thus Fig. 16 shows the Mean Distance to the Origin for all agents, the Mean Distance to the Closest Neighboring agent MDCN, and the Mean Distance of an agent to all other Agents, with standard deviations, all as a function of the Attribute Influence Bound AIB. Fig. 16 shows a complex structure for AIB below 70, which is the region in which agents are coalescing to form groups, and is the region with the greatest statistical variability as the larger standard deviations show. In Fig. 16b, a notable feature is seen for AIB above 80: the linear upper curve shows the coalescence of the curves for the Mean Distance and the Mean Distance to Closest Neighbor MDCN at about AIB$=70$, and that both increase linearly with AIB.  The merging arises primarily because only 2 and sometimes 3 agents remain so all agents are neighbors. As we have discussed, the final stages in simulations with these large values of AIB are dominated by repulsive forces, and the agents repel each other until they are AIB apart and they can no longer communicate.  This behavior is also reflected in the blue lower curve in Fig. 16b in which the Mean Distance to the Origin increases linearly and equals one half of AIB for AIB$> 70$ because agents tend to be on opposite sides of the origin.   

The behavior is clarified in Fig. 17, which shows the Mean Distance to Closest Neighbor divided by AIB (MDCN/AIB), as a function of AIB. For AIB above about 70 or 80, where repulsive forces begin to dominate the end point, the ratio is precisely 1, clearly indicating the final 2 or 3 groups are separated by AIB.  For AIB from 1 to 20, where attractive forces dominate the end point, the ratio decreases rapidly from 7 to about 1.4, where it remains until about AIB=35, when it begins to drop to 1. The overall behavior of the MDCN is not in agreement with the '2R conjecture', which suggests that the separation between nearest agents would be about $2\times$ AIB for an opinion averaging model\cite{blondel}. The disagreement is a consequence of the $1/r^{2}$ force law in the FVM and not solely a consequence of the presence of repulsive forces since the disagreement occurs for AIB less than 40.  

\begin{figure}
\includegraphics[scale=0.6]{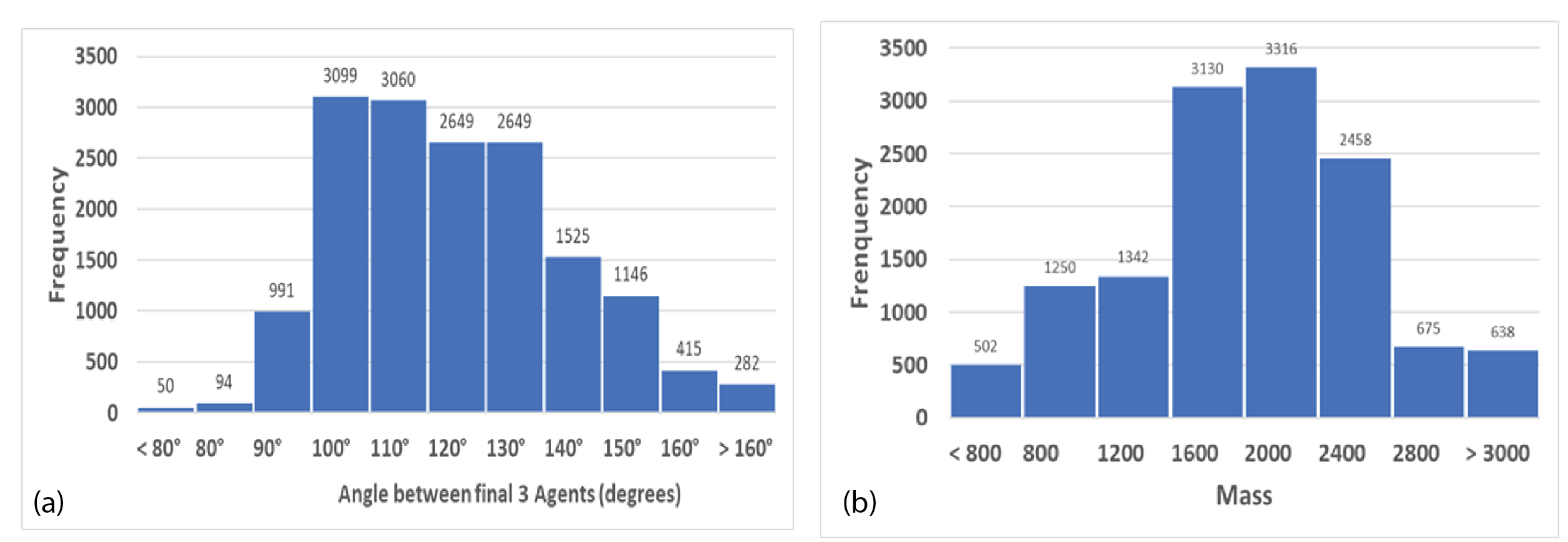} 
\caption{Histograms of angle and mass for final states with three agents: a) Angle between adjacent agents. Mean=120$\degree$, SD=17.5 $\degree$. b) Mass of final agents. Mean=2000, SD=653 3 final agents.}
\label{fig15}
\end{figure}


\begin{figure}[ht]
\hspace{10pt}
\includegraphics[scale=.60]{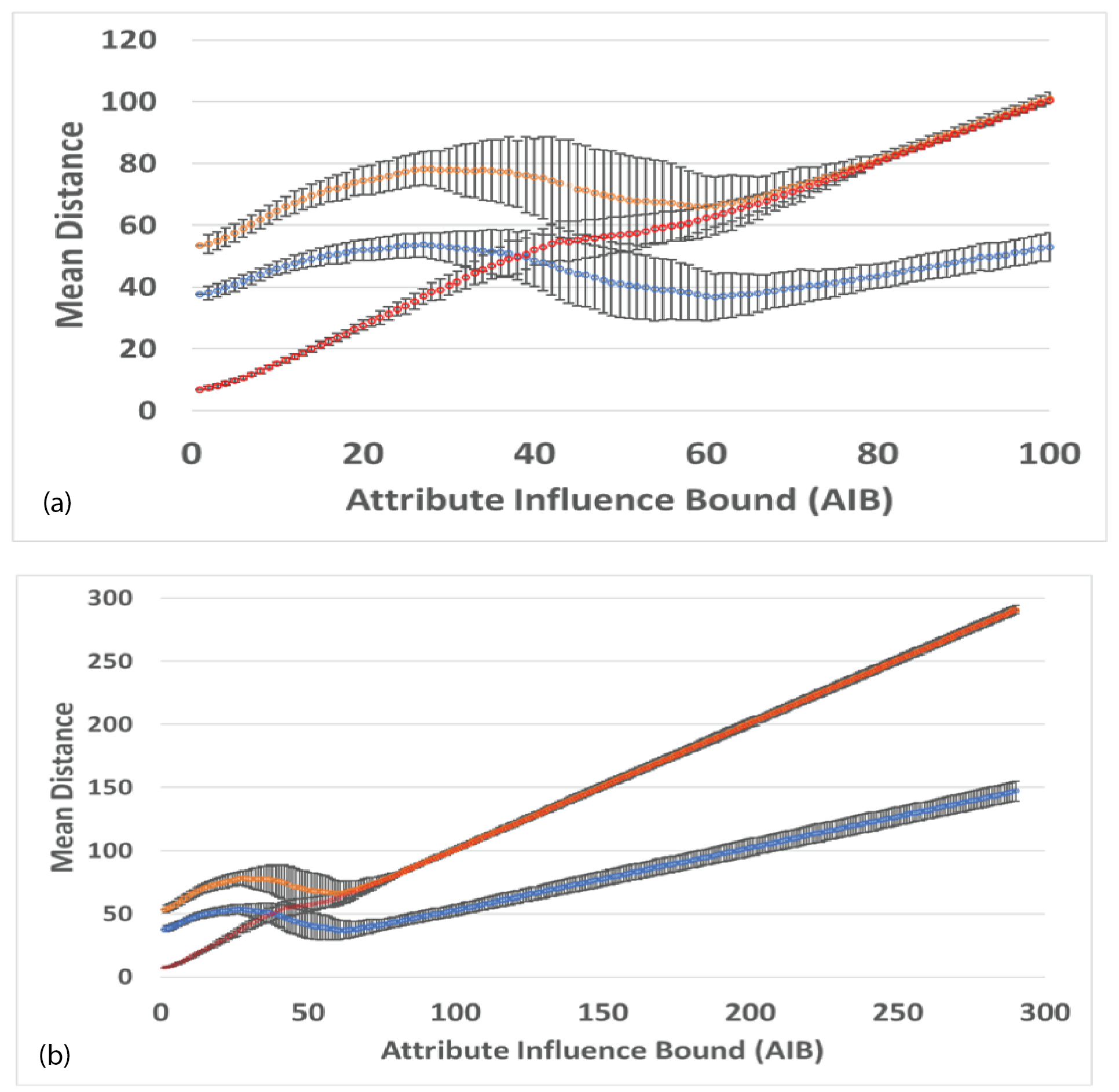} 
\caption{Mean Distances at the end of a simulation with Standard Deviations. Orange=Mean Distance between Agents; Blue=Mean Distance to Origin; Red=MDCN Mean Distance to Closest Neighbor. a) AIB from 0 to 100. b) AIB from 0 to 280. }
\label{fig16}
\end{figure}

\begin{figure}
\hspace{10pt}
\includegraphics[scale=0.45]{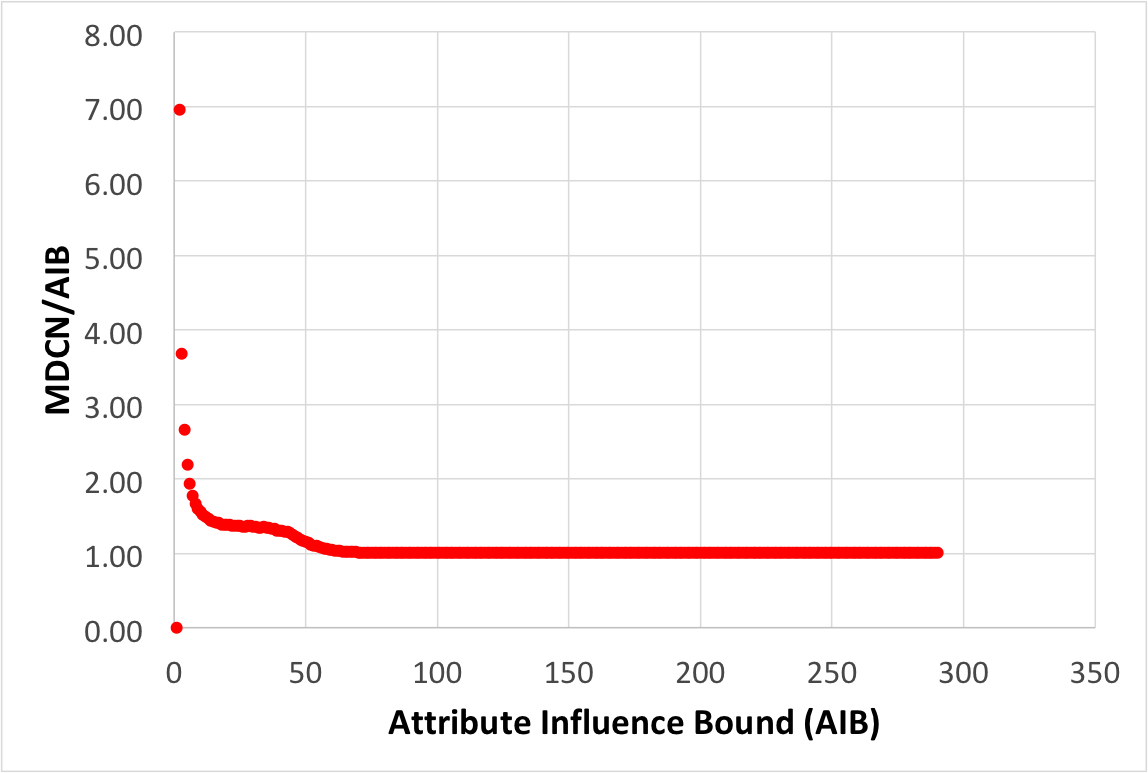} 
\caption{The ratio MDCN/AIB = Mean Distance to Closest Neighbor/AIB as a function of AIB, Standard Conditions.}
\label{fig17}
\end{figure} 

The standard deviations shown in Figs. 13 and 16 are essentially the same whether 100 or 1000 runs are averaged per point. Although we have a deterministic model, there is significant variability in individual simulations due to the uncontrollable variations in the normal initial distributions.

\subsection*{Role of repulsive forces}
There is an option in the model, to set all repulsive forces to zero. In order to illustrate the role of repulsive forces, we did a series of identical simulations with and without repulsive forces.  For every simulation with a different seed, the detailed role of repulsive forces can be different but generalities appear.  
Fig. 18 shows a representative overlay of three simulations with AIB 30, 60, 180 done with and without repulsive forces: Fig. 18a shows that the role of repulsive forces for AIB=30 is minor; Fig. 18b shows that for AIB=60 repulsive forces are beginning to change the evolution of the simulation and the final location of agents; Fig. 18c for AIB=180 shows that repulsive forces play an important role both in the evolution of simulation and especially in the last phase. 

\begin{figure}[h]
\includegraphics[scale=0.68]{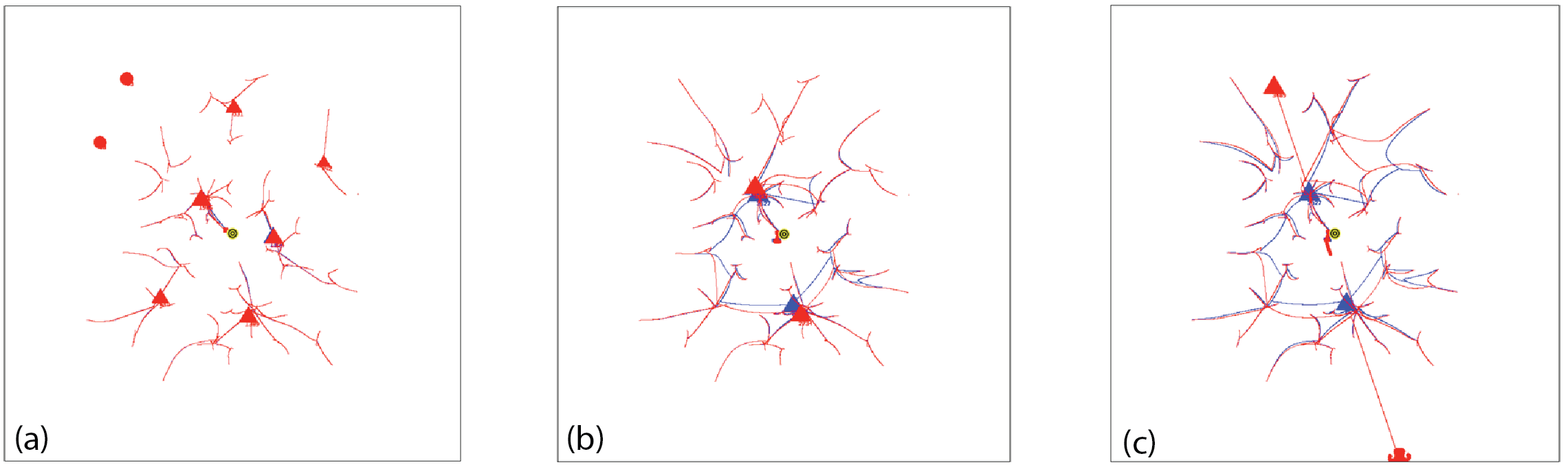}
\vspace{2pt}
\caption{Overlays of simulations with (in red) and without (in blue) repulsive forces. Seed = 91, Standard Conditions. a) AIB=30; b) AIB=60; c) AIB= 180.}
\label{fig18}
\end{figure}

\subsection*{Initial conditions leading to consensus} 

By consensus we mean the final state of a simulation in which all initial agents are in one group. From a mathematical perspective, a sufficient  condition for consensus is locating all initial agents inside one quadrant and having AIB large enough so all agents can communicate.  If all agents are inside one quadrant, then the largest angle between any two agents is less than $90\degree $ so all the forces between agents are attractive.  If all agents can communicate, then they can respond to attractive forces and form groups, which coalesce eventually into one group demonstrating consensus. We can generalize the condition that ensures consensus by simply requiring that all initial agents subtend an angle of less than $90\degree $.

This requirement for consensus can be met by locating the mean of a normal initial spatial distribution of agents away from the origin, and having a standard deviation that is small enough. However, often distributions have outliers that lead to larger angles between agents and repulsive forces, so the final group may not include all agents.  

We have done 44,100 simulations exploring conditions that lead to consensus or near consensus, in which at least 90$\% $ of total mass is concentrated in one group. In our simulations, the mass of the final group is a proxy for the number of agents that compose the group.     

We varied the center of the initial distribution from the origin up to 50 units in every direction in steps of 5, thereby exploring a $100 \times 100$ space centered on the origin.  Specifically, we let the center $(X_{mean}, Y_{mean})$ start at (-50, -50) then increased each coordinate by multiples of 5, so (-45, -50), (-40, -50)  etc, then increased from (0,0) by multiples of 5 so we had (5, 0) up to (50,50), with all combinations of multiples of 5 in each coordinate, for all quadrants. Except for the change in the center of the normal distribution, Standard Conditions applied, SD=30 and AIB=180.  For each location $(X_{mean}, Y_{mean})$ in this $21 \times 21$ matrix of initial locations, 100 simulations were conducted.

There was total consensus, with one final group, for $3.8\%$ of the simulations; for $66.4\%$ of the simulations, there were two final groups and one had over $90\%$ of the agents. 

Fig. 19a shows the probability for a simulation centered on $(X_{mean},Y_{mean})$ to result in total consensus, and Fig 19b, for $\geq 90\% $agreement.

\begin{figure}[ht]
\includegraphics[scale=0.6]{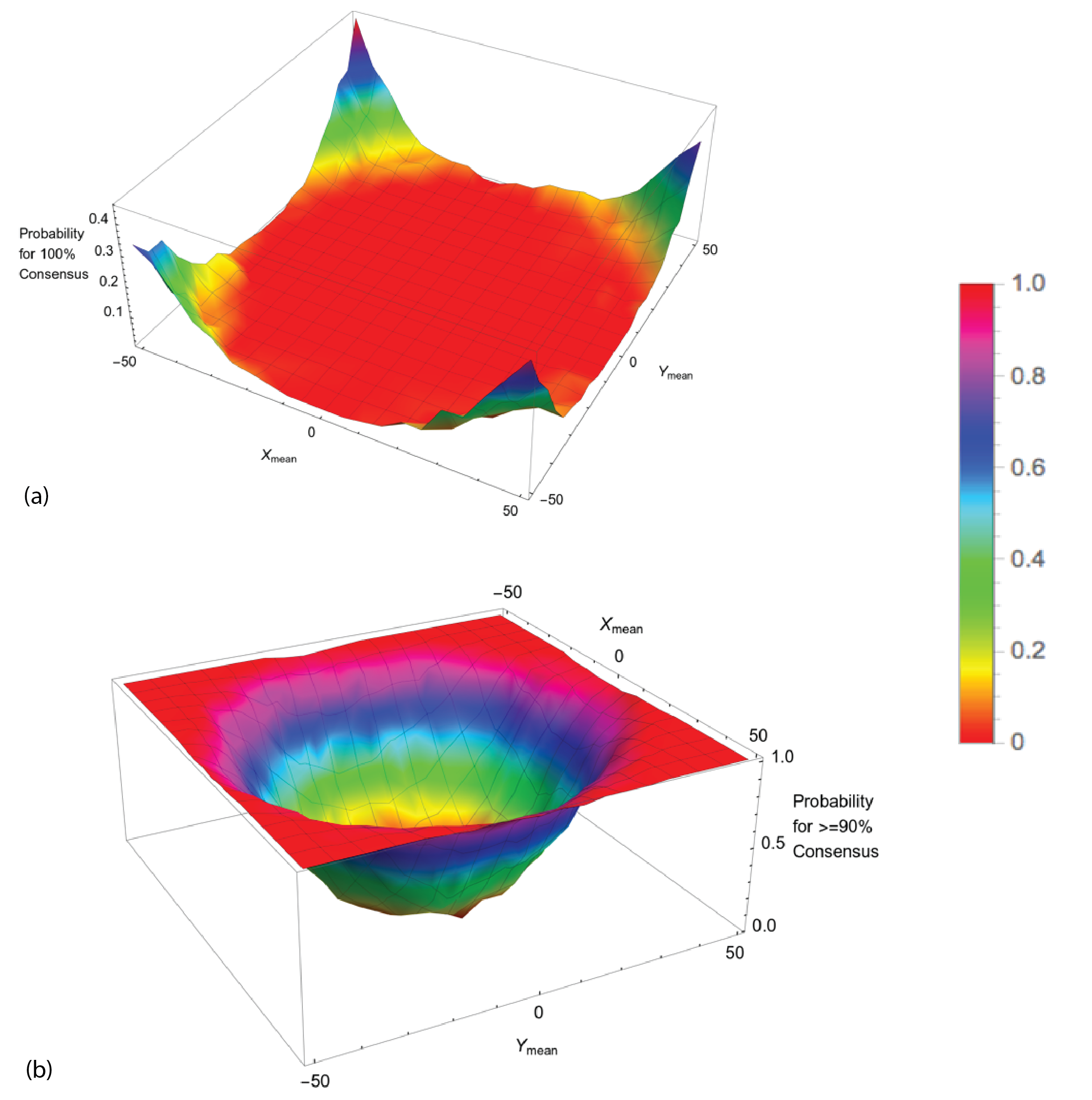}
\caption{Effect of the location of the center of the initial distribution $(X_{mean},Y_{mean})$ on the degree of agreement present at the end of the simulation: a) For consensus (100\% agreement); b) For agreement of $\ge $90\% of agents at end of simulation.}

\label{fig19}
\end{figure}

The statistical data show that consensus occurs about half the time and near consensus ($\ge$90\% agreement) over 95\% of the time for initial attribute distributions that are centered at a point at least 50 units from the origin, for example at (35,35). The location of the final large mass group tends to be 5 to 20 units more distant from the origin than the initial center of the distribution.

There are, however, numerous lower probability occasions when total consensus is achieved when the initial distribution is much closer to the origin, but just happens to have the precise arrangement needed.  An example is shown in Fig. 20. The initial arrangement centered on (20,0) is shown in Fig. 20a, and has agents encircling the origin;  the final state of consensus is shown in Fig. 20b. This simulation can be seen at \url{ https://www.youtube.com/watch?v=MCEbhAqkbjI}.        
In this simulation, agents within each quadrant first merge into groups and then these groups merge. Groups in opposite quadrants (first and third, or second and fourth) cannot merge directly because only repulsive forces are present. The formation of consensus in these cases is a delicate balance. Adding or removing one agent can upset this balance.

\begin{figure}[ht]
\hspace{15pt}
\includegraphics[scale=0.6]{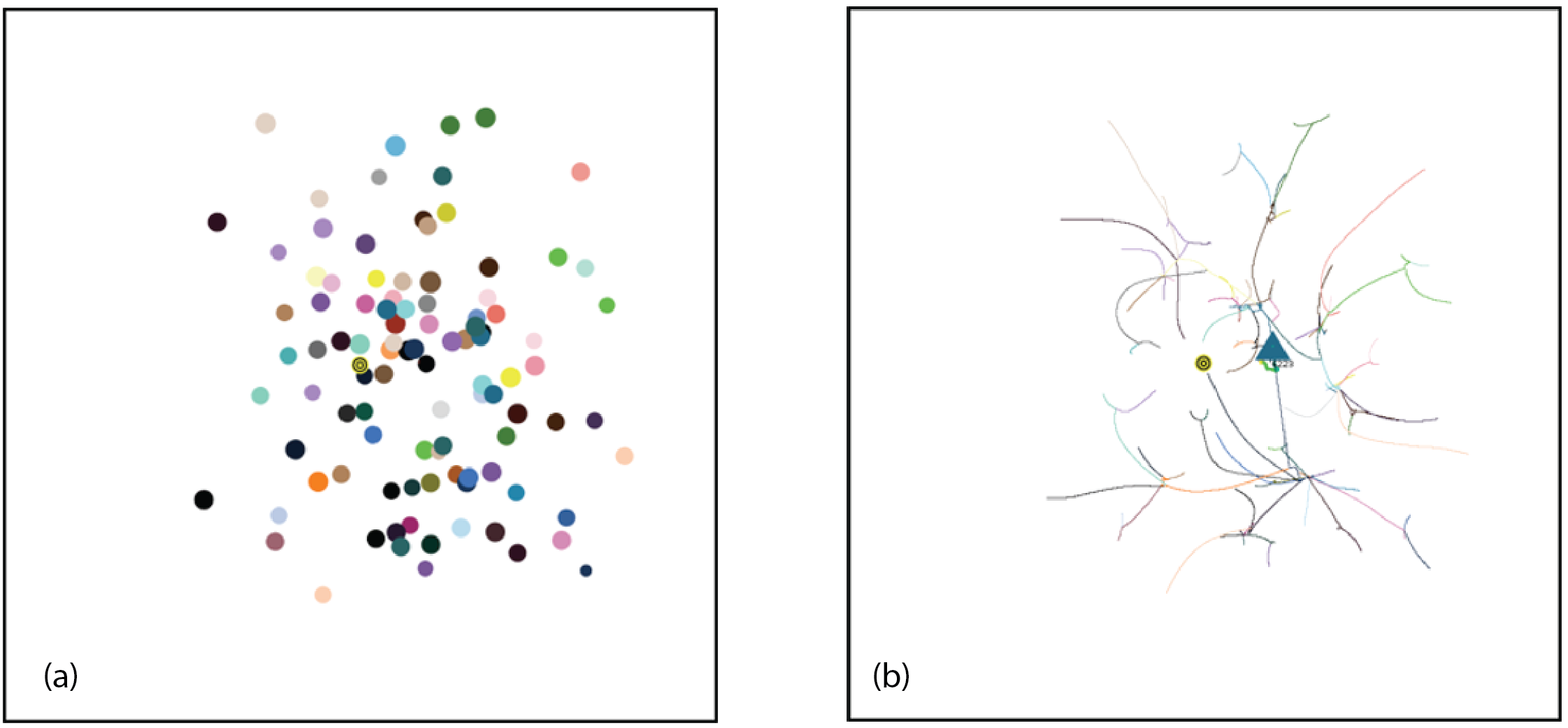}
\vspace{5pt}

\caption{Example of random distribution centered at (20,0) leading to consensus. Seed= 91, AIB=60, final location of group =(19.6, 4.9): a) Start and b) End of simulation.}

\label{fig20}
\end{figure}

As an aside, if the center of the initial normal distribution is moved a small amount, it can affect the final outcome of the simulation. For AIB=100, moving the center of the distribution from (0,0) to (5,5) led to small changes in the final state, but moving it to (10,10) led, in some cases, to significant changes in the angular location of the final few agents. 

\subsection*{Effect of adding an individual agent}
We have also done a preliminary exploration of the role of adding one additional agent to the initial normal distribution. The effect depends very much on the particular geometry, the value of the AIB, and the mass and precise location of the additional agent. We have not analyzed the effects of the additional agent during the simulation, just the effect on the final state. To illustrate, we added one agent at (0,30) on the y-axis to the same simulations as shown in Figs. 8-11.  For AIB=30, an additional agent of mass 500 to 1000 led to the appearance of an additional group near (0,30) that essentially took half the mass of each of the neighboring groups on the right and left that were within the AIB (Fig. 21a). No other changes resulted because of the small AIB.  On the other hand for AIB=180, adding an agent of mass of only 100 at the same location caused the final state to change from three groups (Fig. 11b) to just two groups (Fig. 21b).  Why did this major change occur?  The addition of the single agent between the top two groups shown in Fig. 11b essentially attracted each group slightly to the center line, reducing the angle between them from over 90\degree to less than 90\degree, causing the top two groups to be attracted to each other and ultimately merge.  If the additional agent was added near the location of one of the three final agents, it did not have such a significant effect, even if the mass was 1000. 
\begin{figure}[ht]
\hspace{15pt}
\includegraphics[scale=0.6]{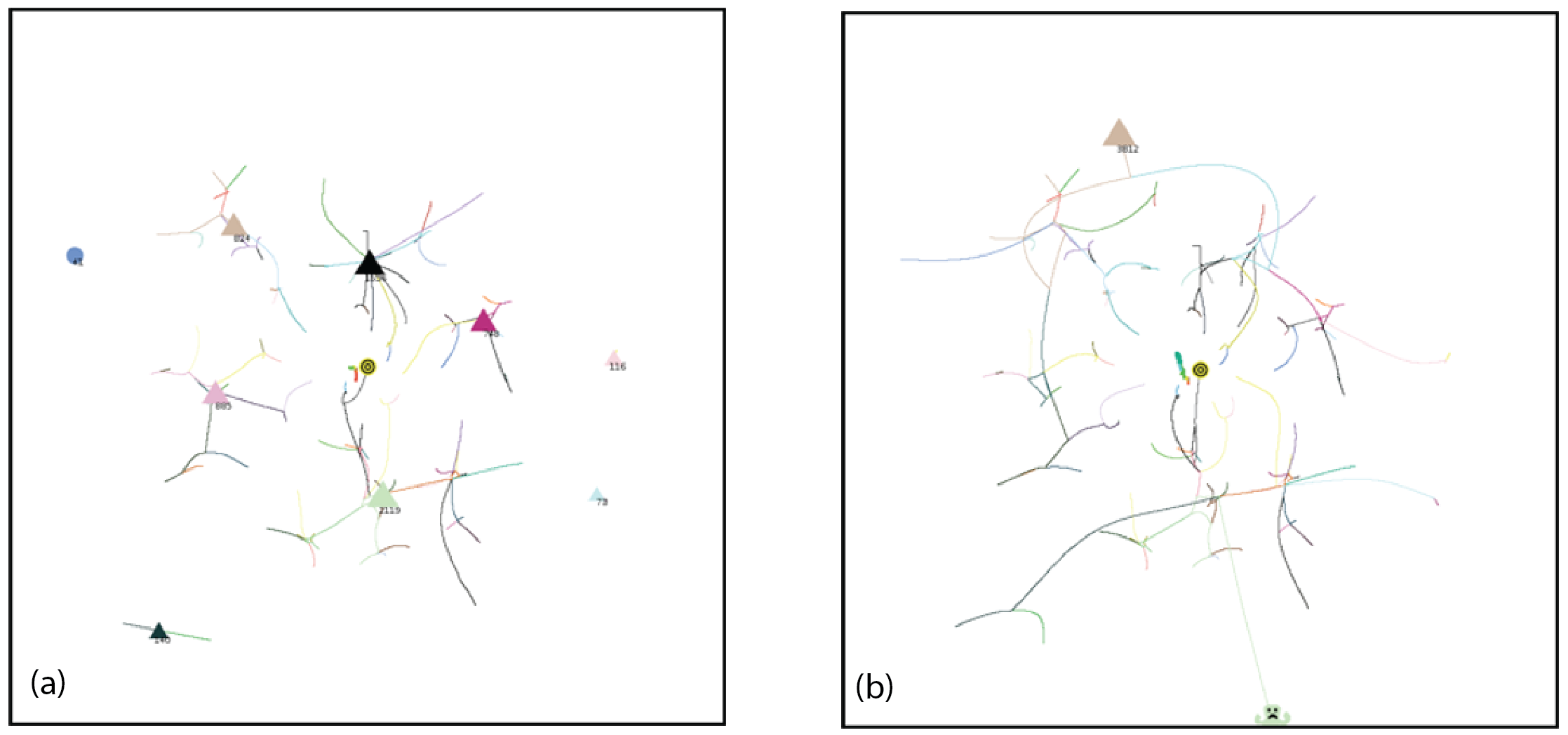} 
\vspace{5pt}
\caption{Effect on final state of adding one agent to the initial configuration. a) AIB= 30, agent of mass 500 added at (0,30). Compare to Fig. 11a. b) AIB=180, agent of mass 100 added at (0,30). Compare to Fig. 11b.}
\end{figure}

\subsection*{Watersheds or attribute sheds}
We noticed an analogy between the behavior of rain falling on a hilly topology and the movements of agents during a simulation.  On a hilly surface, rain flows one way or the other according to watersheds that divide the space. In analogy, we observe divisions in attribute space that agents do not cross during a simulation. We might call these divisions attribute sheds.  Fig. 22 shows the divisions visible in the endpoint for a simulation with AIB=200. The continuous formation of attribute sheds is observed during an entire simulation. Often agents within a region bounded by attribute sheds merge into a single group, and adjacent attribute sheds may merge as the simulation progresses. No agents cross these boundaries during the simulation.  We hypothesize that the watersheds present at the end of a simulation will always intersect the origin if the initial distribution is centered on the origin.

One approach to understand watersheds or attribute sheds is in analogy with the N-body problem in gravitational physics (which is where the FVM began): introduce a potential, compute equipotential curves, and then the forces would be normal to these curves.  In our situation, the potential difference of interest would correspond to the amount of work it would take to move agent A at its current location to within the Coalescence Radius CR of agent B.  If there is an attractive force between them, then the work would be negative; if the force is repulsive it would be positive.  Unfortunately, calculations show that for the force law in the FVM the amount of work done depends on the path taken from A to B.  This means that a regular potential function which gives a potential as a function of position alone cannot characterize the arrangement of agents.   Another approach is based on the spacing between the initial arrangement of agents.  Typically the widest path with no agents will become, at some point in the simulation, a boundary of a attribute shed.  Certainly, the information in the initial arrangement of agents is sufficient to predict the outcome of the simulation, however, we have not found an effective approach using this initial arrangement to predict the attribute sheds.  Running the simulation may be the most practical way of determining the precise arrangement of attribute sheds evolving during the simulation.

We can analyze the general process of the formation of attribute sheds and how it relates to the inverse square force law.
As a consequence of the inverse square law, there is a general process taking place during the simulation:\medskip

1.  At the beginning of the simulation, the agents closest to each other experience the strongest forces of attraction and move towards each other and form a group.  All agents close enough to each other participate in this coalescence.  All agents within the group that is formed have been acquired from the region surrounding the final location of the group, so the group represents consensus within this region, and is surrounded by a boundary of some width that includes no agents.  This boundary can be considered a watershed or attribute shed at this point of the simulation.  No other agents have crossed this attribute shed during this phase of coalescence, otherwise they would have been drawn into the group.\medskip

2. Next the groups that are now closest to each other or to other agents are attracted to each other and move towards each other.  The initial mean distance between these groups will be a bit larger than in the first coalescence because of the inverse square law, but essentially the same process will take place at a slightly larger scale.   The groups will move towards each other and form a new group.  All agents within a certain distance will coalesce, thus this new group will represent consensus in the larger region and will be surrounded by a new boundary that has no agents, forming a new attribute shed.  This attribute shed may intersect with previous attribute sheds since the movement of agents and groups has changed the landscape of forces.\medskip

3. This process will continue until the end of the simulation.  The groups remaining at the end of the final coalescence will be the final agents and will be surrounded by new attribute sheds. \medskip 

4. For simulations with a large AIB, the final agents within each attribute shed may undergo a repulsive phase in which they cross trajectories, but we can still see the distinct attribute sheds that divide the attribute space into distinct regions.\medskip
 
The attribute sheds can be interpreted as representing an effective or local equipotential curve. The energy required for each member of the group that formed to move from about their initial location near the edge of the attribute shed to the final location of the group, is about the same, and in that sense the attribute shed is like an equipotential surface.  It is not an equipotential surface in the conventional sense because the change in energy depends on path.  If it were a true equipotential curve, the forces and trajectories would be normal to it, however, this is not always the case in this system.

\begin{figure}
    \hspace{1pt}
    \includegraphics[scale=0.125]{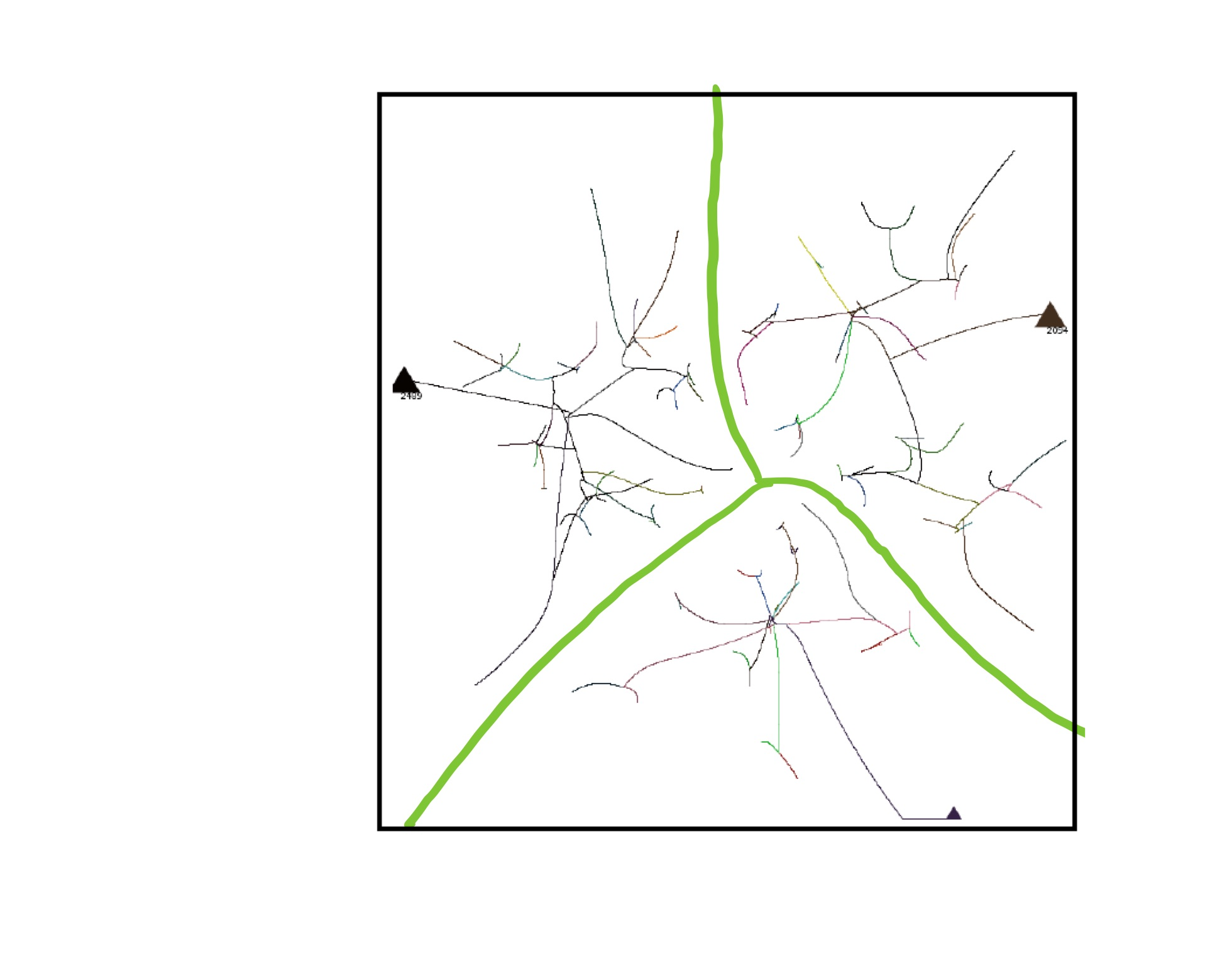}
    \caption{Attribute sheds present in a simulation with AIB=200, shown with thick solid lines. Agent trails show that no agents have crossed these dividing lines during the simulation.}
    \label{fig22}
\end{figure} 

\subsection*{Effect of dimensionality on simulations}
The FVM can be easily generalized to higher dimensions.  We did several simulations using Standard Conditions, in which we varied the dimensionality of the vector space from 3 to 15, with AIB values that spanned the spaces. In general, when the simulation ended, there were only 2-3 groups remaining in very different regions of the higher dimensional attribute space, suggesting that for the limited conditions tested the presence of polarization in a connected universe is independent of the dimension of attribute space. 

\section*{Discussion}

The FVM is a deterministic model yet many of the outcomes show strong statistical variations due to the statistical nature of the initial agent distributions.  The relationship between deterministic and stochastic models and initial conditions is worth discussing briefly. 

One system illustrating the role of random initial conditions in deterministic equations is seen in Brownian/Langevin dynamics\cite{kampen}.  There are two forms of randomness: the sampling of an initial condition and the Brownian noise.  Reducing the temperature to zero removes the Brownian noise, leading to deterministic equations but with random initial conditions, akin to the FVM. This system also has a link to quantum theory: the Langevin equation can be derived using path integral methods characteristic of quantum theory\cite{janssen}. 

There is an interpretation of quantum mechanics due to David Bohm that is completely deterministic, as is our model,  yet his interpretation agrees with the statistical predictions that are fundamental to the standard Copenhagen interpretation of quantum theory. In Bohmian mechanics, the observed variability in experimental results is due ultimately to the inability to define the initial conditions precisely enough \cite{bohm}.  We have a similar situation with our deterministic model in which a statistically defined normal distribution inherently has sufficient variability so that our deterministic model gives a random polarization outcome\cite{atom}.

Is the behavior of agents in any way like the behavior of atoms or atomic structures? Fortunato and Castellano observe: "The idea that humans, in particular circumstances, behave as \textit{social atoms}, i.e. they might obey rules that yield predictable collective patterns, just like atoms and molecules, dates back many years and has several illustrious advocates including political philosopher John Stuart Mill and social scientists Auguste Comte and Adolphe Quetelet. The principle is that, when a great multitude of individuals interact, the choice of each of them, which in principle is free, is constrained by the presence of the others, and regular collective behaviors may result" \cite{Fortunato_2012}\cite{Buchanan_2007}.

It is in this spirit that we model the interaction of agents in the FVM on the general behavior of electric dipoles. Why an electric dipole? A dipole is composed of two charges with opposite signs that are separated by a small distance. A dipole is represented as a vector, which indicates its magnitude and its direction, corresponding, in principle, to the way we represent agents in our model. A characteristic electric field surrounds the dipole. The way a dipole interacts with other dipoles depends on the relationship between the corresponding electric fields which are indicated by the vectors, how close they are and how they point. Dipoles can attract each other, repel each other and rotate each other. To describe forces between real dipoles leads to equations that are complex.  We represent agents with vectors, but simplify the mathematics in a way that hopefully captures the relevant fundamental behaviors.

As Castellano et al state: "In most situations, qualitative (and even some quantitative) properties of large-scale phenomena do not depend on the microscopic details of the process.  Only higher level features, such as symmetries, dimensionality, or conservation laws, are relevant for the global behavior.  With this concept of universality in mind, one can approach the modelization of social systems, trying to include only the simplest and most important properties of single individuals and looking for qualitative features exhibited by models \cite{castellano}."

\section*{Conclusion}

We have introduced the Force Vector Model of social influence, compared it to the HK bounded confidence model, explaining similarities and differences.  For example, both models are deterministic and both have criteria for the formation of groups or opinion clusters, both are characterized by a confidence bound, both can be interpreted as having forces that depend on the distance between agents in attribute space:  a force directly proportional to the distance for the BC models and a force inversely proportional to the distance squared for the FVM.  Bounded confidence models can be modified to have repulsive forces that are based on the distance between agents, whereas for the FVM, the repulsive forces are a consequence of the Attribute Orientation AO or relative direction of the vectors for the agents. The different force laws produce distinctly different simulations. The FVM is designed as a simulation of a physical system, providing a smooth and continuous evolution of agent attributes, in which the inverse square force law results in the progressive coalescence of agents that are closest to each other.  

The FVM is an integrated model in which agents are represented by vectors in the visible 200 x 200 attribute space. During most simulations agents remain in the visible space, however, the agents can move beyond these boundaries without limit.  The behavior of agents is determined by two simple vector equations, one for the force between agents and one for the change in an agent’s vector due to the force. The vector equations are readily generalizable to higher dimensions.  The appearance of attractive and repulsive forces between agents and the effect of the forces on the agents is based on these two equations, and the AIB which limits communication to those agents within the AIB.   

The movements of agents are recorded during the simulation, along with a variety of other metrics.  In each time iteration, the maximum movement of an agent is limited by adaptive time stepping to a small preset value, the Coalescence Radius, ensuring a smooth continuous trajectory as is expected in the simulation of a physical system. We verified that adaptive time stepping does not change the outcomes of simulations and the simulations are stable with respect to variations in the model parameters.  In addition, we verified that there was no inherent angular bias on the outcomes of simulations.  For a specific simulation, however, the location of the final groups depends on the fine details of the initial normal distribution, and changing just one agent can, in some cases, affect the outcome.   

Over 120,000 simulations were completed.  The outcomes of simulations depend on the initial conditions, which includes the values of the model parameters, the initial spatial arrangement of agents and their masses, and on the value of the Attribute Influence Bound. Most of the simulations were done using our Standard Conditions, in which we use the default values of model parameters, a symmetric normal distribution of 100 agents about the origin with a standard deviation for x and y both equal to 30 units, and a normal distribution of masses with mean of 60 and standard deviation of 15, with equal active and passive masses.  For this set of Standard Conditions, a baseline description of outcomes was presented.  

A broad spectrum of behaviors was observed and all simulations were different in their details but numerous statistical regularities appeared. For example, the Center of Attributes, which is the mass weighted average of the attributes for all agents is approximately constant during a simulation. There is an exponential decline in the number N of final agents from 98.6 to 3.2 as the AIB is increased from 1 to 52.  The formation of these groups is primarily a consequence of attractive forces.

For a small AIB$ \le 40$, numerous groups distributed within attribute space remain at the end point of the simulation, and the separation of an agent to its closest neighboring agent (MDCN) is typically a little greater than AIB. For example, for an AIB of 30, about 9 agents remain, about 7 of them groups, with a MDCN of 40 or 1.3 AIB. The attributes of each group are essentially the average attributes of all its members, reflecting consensus within a region of attribute space. Repulsive forces are less than 1\% of the total final force. 

For AIB of 40-60, we see coalescence of the groups closest to each other, producing 8 to 3 groups with separations about equal to  AIB.  For these conditions, the repulsive forces are generally less than $20-50\%$ of the total final force, so the separations between groups tend to remain about the same if repulsive forces are turned off during the simulation. This condition arises because repulsive forces can only be present between agents in different quadrants, which means they tend to be farther apart. However, if AIB is not large enough, these agents cannot communicate.

If the AIB is larger, say 60-100, we typically find end points in which 4 to 2 groups tend to be arranged approximately symmetrically about the origin: four agents in a square, three in an equilateral triangle, or two on opposite sides of the origin, separated from each other by about AIB so MDCN=AIB, and the mean distance to the origin is about 1/2 AIB.  The masses of the N final groups appear to be distributed approximately normally with a mean of 100x60/N and a SD of about 500. At the end of a simulation, repulsive forces are very significant, representing 70-100\% of the total final force. 

In a typical example for AIB between 80 and 100, we may have two remaining groups at the end of the simulation, and each has a roughly comparable fraction of the total number of agents. The attributes of each group are essentially the average of the attributes of all its members, which shows consensus within this segment of the population. The members of each group were initially within approximately a $+90 \degree$ to $-90\degree$ interval about the origin, and they merged over time into one group.  We note therefore that agents with attribute orientations AO differing from each other by up to almost $180 \degree $ may be within this group, which means there is consensus over a diverse population, reflecting a significant change in attributes during the simulation.  Nevertheless the Center of Attributes is approximately constant.

For AIB over about 100 the mean number of final agents is about 2 to 2.25, and typically a final phase of the simulation occurs in which 100\%  repulsive forces drive the groups farther apart, until the separation between each is approximately AIB.  If the AIB is large enough, the groups will be driven to the walls or corners of the visible universe or well beyond. In this final phase the group attributes become more extreme than the initial opinions of its members, leading to group polarization.  

The typical arrangement of groups in the late stages of a simulation is reminiscent of the Rule of Three in economics, which predicts that there are typically three major competitors in a given industry in a free market \cite{ivy}.  Research has verified this general rule \cite{three}.

Another way to view the polarized results of the simulations is in terms of mimetic desire as proposed by the anthropologist Rene Girard \cite{renebook}. Desire, he maintains, has its fundamental origins in terms of our perception of what others, whom we model, desire. All conflict, competition, and rivalry originate in mimetic desire, which eventually can lead to destructive stages of conflict between individuals and social groups that require them to blame someone or something in order to defuse conflict through the scapegoat mechanism. 

To determine the conditions under which consensus is obtained, we did 40,000 simulations in which we systematically varied the center of the initial distribution of agents within a square 100 units x 100 units centered on the origin.  We find that if the center of the initial distribution of agents is at least 50 units from the origin, for example at (35,35), then consensus occurs in about 50\% of the simulations, and near consensus, meaning 90\% agreement, occurs over 95\% of the time.  In addition, there were individual simulations ending in consensus when the center of the symmetric normal distribution was only 15 or 20 units from the origin.

To the extent that the FVM model does reflect human behaviors, we can hypothesize about the cognitive and emotional processes that might occur within the agents.  Probably a key factor active in the initial coalescence phase would be an agent's desire to not be different, to fit in, to feel connection, to feel unity and harmony, to cooperate.   These are human traits that have driven the evolution and socialization of our species \cite{harari}. The phase of joining with another agent to form a group may reflect the drive for security, for a sense of power and togetherness by being a member of a group.

As the group increases in size, a stronger sense of tribal or ethnocentric identity, including groupthink, may develop. In the simulations, this is the time in which the distance between an agent's initial attributes and the group attributes may begin to increase.  This is the time when agents may begin to base their personal identity on embracing the group attributes, and the group psychology begins to assume an independence. Other groups have formed with different attributes. 

Initially, the groups tend to coexist in different regions of attribute space, and groups that are not too different may combine with each other to form larger groups, with stronger in-group identities. An agent's identification with the group becomes more solidified, and, as repulsive forces begin to dominate the simulations, the groups continue the development of their group attributes or identities in contrast to each other, in competition with each other, with the threat of each other. Communication between the groups only drives them farther apart. Extreme polarization results.

From another perspective, perhaps it is appropriate to consider the behavior described by the model as driven by what has been referred to as the limbic system or the paleomammalian cortex in which behavior is based on primal, gut level responses.  Individuals may rely on this approach of "primitive automaticity" in an effort to cut through the complex information/misinformation landscape \cite {cialdini}.  That a simple deterministic model appears to capture some of the complexities of observed behavior may even lead to the question of whether free will is involved in the choices agents make during the evolution of a simulation.

As we have mentioned, the baseline calibration of the FVM was based on Standard Conditions and primarily considering endpoints in the simulations. We need to explore the model beyond these initial conditions, and consider, for example, simulations in which the initial x and y distributions are not the same so there would be a preference for final groups at certain angles.  We need to consider the possibility of having two independent distributions of initial agents. We have also assumed passive and active masses were equal. We need to explore the use of the passive mass to simulate, for example, resistance to the influence of other agents, which has been called stubbornness or apathy in political systems\cite{hegselmann2}. More characterization of the time dependence of the simulations would help in understanding the phases of development in simulations and perhaps lead to the development of feedback loops in the FVM. 

Some modifications to the FVM might increase its capability to model the organization of social systems, including, for example, more nuanced treatment of groups, including a life cycle, fracturing of groups, and different types of groups that may obey different laws.  Having an Attribute Influence Bound AIB that can vary in time might be of value. In evolution, tribal and small group stages last much longer than empire stages.  Another modification might include a metric that is not Euclidean, but treats some dimensions differently from others or reflects curvature in the universe of attributes. 

As Flache et al noted in their comprehensive review, “Despite much research, social influence remains one of the most puzzling social phenomena \cite {flache}.  
One important path to progress is to compare models to historical data.  Our hope is to develop collaborations and apply the FVM to historical data and let this critical feedback guide the further development of the model.

\section*{Supporting information}



\paragraph*{S1 Datafile.}
 The file FVMsimsAIB1to290.csv
 includes the NetLogo results of 2900 simulations with Standard Conditions, for AIB from 1 to 290 in steps of 1, 10 simulations with 10 different seeds were done for step.





\section*{Acknowledgement}
 We would like to extend our appreciation to Mary Maclay for analyzing and graphing data so carefully, and for her many useful comments and observations.  
\nolinenumbers

%
%
%



\begin{thebibliography}{10}
\bibitem{other}
We have not done any detailed investigations to validate the following
  possibilities, but consider that some possible applications of the model may
  include: opinion dynamics, the coalescence of car manufacturers from many to
  a few that compete with each other; the evolution of social attitudes; the
  formation of families, communities, city states from a collection of
  individuals; the development of political parties and political or economic
  polarization; groupthink; the creation of social groups with common
  interests; the evolution of cultural behaviors; the recruitment of
  individuals for various missions; and the recruitment of cells in collective
  brain activity.
\bibitem{atom}
It is striking that Fig. 3a suggests a semiclassical limit of the hydrogen
  atom, in which each dot represents a measurement of the location of the
  electron orbiting around the nucleus in an excited state. The approximate
  diameter of the orbit is determined by the value of AIB in the FVM and by the
  principal quantum number in quantum mechanics. Each measurement would knock
  the electron out of the atom from a random location, so a series of
  measurements would be needed to determine all the dots and the approximate
  trajectory of the bound electron.
\end{thebibliography}

\end{document}